\documentclass[aps,prc,twocolumn,showpacs,superscriptaddress,groupedaddress]{revtex4}  
\usepackage{graphicx}  
\usepackage{dcolumn}   
\usepackage{bm}        
\usepackage{amssymb}   
\usepackage{amsmath}
\usepackage[errorshow]{tracefnt}
\usepackage{csquotes}
\usepackage{longtable}
\usepackage{tabularx}
\usepackage{multirow}
\usepackage{booktabs}
\usepackage{threeparttable} 
\usepackage{rotating}
\usepackage{epigraph} 
\usepackage{xcolor}

\begin{document}


\title{Shape coexistence in the neutron-deficient $^{188}$Hg \\investigated via lifetime measurements}
%
\author{M.~Siciliano}
	\affiliation{INFN, Laboratori Nazionali di Legnaro, Legnaro, Italy.}
	\affiliation{Dipartimento di Fisica e Astronomia, Universit\`a di Padova, Padova, Italy.}
	\affiliation{Irfu/CEA, Universit\'e de Paris-Saclay, Gif-sur-Yvette, France.}
\author{I.~Zanon}
	\affiliation{INFN, Laboratori Nazionali di Legnaro, Legnaro, Italy.}
	\affiliation{Dipartimento di Fisica e Astronomia, Universit\`a di Padova, Padova, Italy.}
	\affiliation{Dipartimento di Fisica e Scienze della Terra, Universit\`a di Ferrara, Ferrara, Italy.}
\author{A.~Goasduff}
	\affiliation{Dipartimento di Fisica e Astronomia, Universit\`a di Padova, Padova, Italy.}
	\affiliation{INFN, Sezione di Padova, Padova, Italy.}
	\affiliation{Wydzia{\l} Fizyki, Uniwersytet Warszawski, Warsaw, Poland.}
\author{P.R.~John}
	\affiliation{Dipartimento di Fisica e Astronomia, Universit\`a di Padova, Padova, Italy.}
	\affiliation{INFN, Sezione di Padova, Padova, Italy.}
	\affiliation{Institut f\"ur Kernphysik der Technischen Universit\"at Darmstadt, Darmstadt, Germany.}
\author{T.R.~Rodr\'iguez}
	\affiliation{Departamento de F\'isica Te\'orica and Centro de Investigaci\'on Avanzada en F\'isica Fundamental Universidad Aut\'onoma de Madrid, Madrid, Spain.}
\author{S.~P\'{e}ru}
	\affiliation{CEA, DAM, DIF, Arpajon, France.}
\author{I.~Deloncle}
	\affiliation{CEA, DAM, DIF, Arpajon, France.}
	\affiliation{IJCLab CNRS/IN2P3, Universit\'{e} de Paris-Saclay, Orsay, France.}		
\author{J. Libert}
	\affiliation{CEA, DAM, DIF, Arpajon, France.}
\author{M.~Zieli\'{n}ska}
	\affiliation{Irfu/CEA, Universit\'e de Paris-Saclay, Gif-sur-Yvette, France.}
\author{D.~Ashad}
	\affiliation{Dipartimento di Fisica and INFN Sezione di Napoli, Napoli, Italy.}
\author{D.~Bazzacco}
	\affiliation{INFN, Sezione di Padova, Padova, Italy.}
\author{G.~Benzoni}
	\affiliation{INFN, Sezione di Milano, Milano, Italy.}
\author{B.~Birkenbach}
	\affiliation{Institut f\"{u}r Kernphysik, Universit\"{a}t zu K\"{o}ln, Cologne, Germany.}
\author{A.~Boso}
	\affiliation{Dipartimento di Fisica e Astronomia, Universit\`a di Padova, Padova, Italy.}
	\affiliation{INFN, Sezione di Padova, Padova, Italy.}
\author{T.~Braunroth}
	\affiliation{Institut f\"{u}r Kernphysik, Universit\"{a}t zu K\"{o}ln, Cologne, Germany.}
\author{M.~Cicerchia}
	\affiliation{INFN, Laboratori Nazionali di Legnaro, Legnaro, Italy.}
	\affiliation{Dipartimento di Fisica e Astronomia, Universit\`a di Padova, Padova, Italy.}
\author{N.~Cieplicka-Ory\'nczak}
	\affiliation{INFN, Sezione di Milano, Milano, Italy.}
	\affiliation{Instytut Fizyki J\c{a}drowej im. Henryka Niewodnicza\'nskiego, Polska Akademia Nauk, Krakow, Poland.}
\author{G.~Colucci}
	\affiliation{Dipartimento di Fisica e Astronomia, Universit\`a di Padova, Padova, Italy.}
	\affiliation{INFN, Sezione di Padova, Padova, Italy.}
	\affiliation{\'Srodowiskowe Laboratorium Ci\c{e}\.zkich Jon\'ow, Uniwersytet Warszawski, Warsaw, Poland.}
\author{F.~Davide}
	\affiliation{Dipartimento di Fisica and INFN Sezione di Napoli, Napoli, Italy.}
\author{G.~de~Angelis}
	\affiliation{INFN, Laboratori Nazionali di Legnaro, Legnaro, Italy.}
\author{B.~de~Canditiis}
	\affiliation{Dipartimento di Fisica and INFN Sezione di Napoli, Napoli, Italy.}
\author{A.~Gadea}
	\affiliation{Instituto de F\'isica Corpuscular, CSIC-Universidad de Valencia, Valencia, Spain.}
\author{L.P.~Gaffney}
	\affiliation{University of the West of Scotland, Paisley, Scotland.}
\author{F.~Galtarossa}
	\affiliation{INFN, Laboratori Nazionali di Legnaro, Legnaro, Italy.}
	\affiliation{Dipartimento di Fisica e Scienze della Terra, Universit\`a di Ferrara, Ferrara, Italy.}
\author{A.~Gozzelino}
	\affiliation{INFN, Laboratori Nazionali di Legnaro, Legnaro, Italy.}
\author{K.~Hady\'nska-Kl\c{e}k}
	\affiliation{INFN, Laboratori Nazionali di Legnaro, Legnaro, Italy.}
	\affiliation{\'Srodowiskowe Laboratorium Ci\c{e}\.zkich Jon\'ow, Uniwersytet Warszawski, Warsaw, Poland.}
\author{G.~Jaworski}
	\affiliation{INFN, Laboratori Nazionali di Legnaro, Legnaro, Italy.}
	\affiliation{\'Srodowiskowe Laboratorium Ci\c{e}\.zkich Jon\'ow, Uniwersytet Warszawski, Warsaw, Poland.}
\author{P.~Koseoglou}
	\affiliation{Institut f\"ur Kernphysik der Technischen Universit\"at Darmstadt, Darmstadt, Germany.}
\author{S.M.~Lenzi}
	\affiliation{Dipartimento di Fisica e Astronomia, Universit\`a di Padova, Padova, Italy.}
	\affiliation{INFN, Sezione di Padova, Padova, Italy.}
\author{B.~Melon}
	\affiliation{Dipartimento di Fisica and INFN Sezione di Firenze, Firenze, Italy.}
\author{R.~Menegazzo}
	\affiliation{INFN, Sezione di Padova, Padova, Italy.}
\author{D.~Mengoni}
	\affiliation{Dipartimento di Fisica e Astronomia, Universit\`a di Padova, Padova, Italy.}
	\affiliation{INFN, Sezione di Padova, Padova, Italy.}
\author{A.~Nannini}
	\affiliation{Dipartimento di Fisica and INFN Sezione di Firenze, Firenze, Italy.}
\author{D.R.~Napoli}
	\affiliation{INFN, Laboratori Nazionali di Legnaro, Legnaro, Italy.}
\author{J.~Pakarinen}
	\affiliation{Fysiikan laitos, Jyv\"askyl\"an Yliopisto, Jyv\"askyl\"a, Finland.}
\author{D.~Quero}
	\affiliation{Dipartimento di Fisica and INFN Sezione di Napoli, Napoli, Italy.}	
\author{P.~Rath}
	\affiliation{Dipartimento di Fisica and INFN Sezione di Napoli, Napoli, Italy.}
\author{F.~Recchia}
	\affiliation{Dipartimento di Fisica e Astronomia, Universit\`a di Padova, Padova, Italy.}
	\affiliation{INFN, Sezione di Padova, Padova, Italy.}
\author{M.~Rocchini}
	\affiliation{Dipartimento di Fisica and INFN Sezione di Firenze, Firenze, Italy.}
\author{D.~Testov}
	\affiliation{Dipartimento di Fisica e Astronomia, Universit\`a di Padova, Padova, Italy.}
	\affiliation{INFN, Sezione di Padova, Padova, Italy.}
	\affiliation{Joint Institute for Nuclear Research, Dubna, Russia.}
\author{J.J.~Valiente-Dob\'on}
	\affiliation{INFN, Laboratori Nazionali di Legnaro, Legnaro, Italy.}
\author{A.~Vogt}
	\affiliation{Institut f\"{u}r Kernphysik, Universit\"{a}t zu K\"{o}ln, Cologne, Germany.}
\author{J.~Wiederhold}
	\affiliation{Institut f\"ur Kernphysik der Technischen Universit\"at Darmstadt, Darmstadt, Germany.}
\author{W.~Witt}
	\affiliation{Institut f\"ur Kernphysik der Technischen Universit\"at Darmstadt, Darmstadt, Germany.}

\vskip 0.25cm


\begin{abstract}
\noindent
\textbf{Background:} Shape coexistence in the $Z \approx 82$ region has been established in mercury, lead and polonium isotopes. 
For even-even mercury isotopes with $100 \leq N \leq 106$ multiple fingerprints of this phenomenon are observed, which seems to be no longer present for $N \geq 110$. 
According to a number of theoretical calculations, shape coexistence is predicted in the $^{188}$Hg isotope. 

\noindent
\textbf{Purpose:} The aim of this work was to measure lifetimes of excited states in $^{188}$Hg to infer their collective properties, such as the deformation. 
Extending the investigation to higher-spin states, which are expected to be less affected by band-mixing effects, can provide additional information on the coexisting structures.

\noindent
\textbf{Methods:} The $^{188}$Hg nucleus was populated using two different fusion-evaporation reactions with two targets, $^{158}$Gd and $^{160}$Gd, and a beam of $^{34}$S provided by the Tandem-ALPI accelerator complex at
the Laboratori Nazionali di Legnaro. 
The channels of interest were selected using the information from the Neutron Wall array, while the $\gamma$ rays
were detected using the GALILEO $\gamma$-ray spectrometer. 
Lifetimes of excited states were determined using the Recoil Distance Doppler-Shift method, employing the dedicated GALILEO plunger device.

\noindent
\textbf{Results:} Lifetimes of the states up to spin $16 \,\hbar$ were measured and the corresponding reduced transition probabilities were calculated. 
Assuming two-band mixing and adopting, as done commonly, the rotational model, the mixing strengths and the deformation parameters of the unperturbed structures were obtained from the experimental results. 
In order to shed light on the nature of the observed configurations in the $^{188}$Hg nucleus, the extracted transition strengths were compared with those resulting from state-of-the-art beyond-mean-field calculations using the symmetry-conserving configuration-mixing approach, limited to axial shapes, and the 5-dimensional collective Hamiltonian, including the triaxial degree of freedom. 

\noindent
\textbf{Conclusions:} The first lifetime measurement for states with spin $\ge$ 6 suggested the presence of an almost spherical structure above the $12_1^+$ isomer and allowed elucidating the structure of the intruder band. 
The comparison of the extracted $B(E2)$ strengths with the two-band mixing model allowed the determination of the ground-state band deformation. 
Both beyond-mean-field calculations predict coexistence of a weakly-deformed band with a strongly prolate-deformed one, characterized by elongation
parameters similar to those obtained experimentally, but the calculated relative position of the bands and their mixing strongly differ. 
\end{abstract}

\maketitle

\section{Introduction}

The regions close to $Z=50$ and $Z=82$ provide unique conditions to study the evolution of nuclear shapes and of collectivity in the vicinity of magic numbers. 
The lead region, in particular, presents a wide range of phenomena related to the nuclear shape, as for instance shape staggering between odd- and even-mass nuclei~\cite{kuhl1977staggering, marsh2018characterization}, shape evolution with mass~\cite{ansari1986shape, john2014shape, john200Pt} and shape coexistence~\cite{andeyev2000triplet, rahkila2010shape}. 
The latter is a characteristic feature of finite many-body quantum systems, such as the atomic nucleus, where structures corresponding to different shapes coexist within a similar excitation energy.

\begin{figure}[h]
\includegraphics[width=0.49\textwidth]{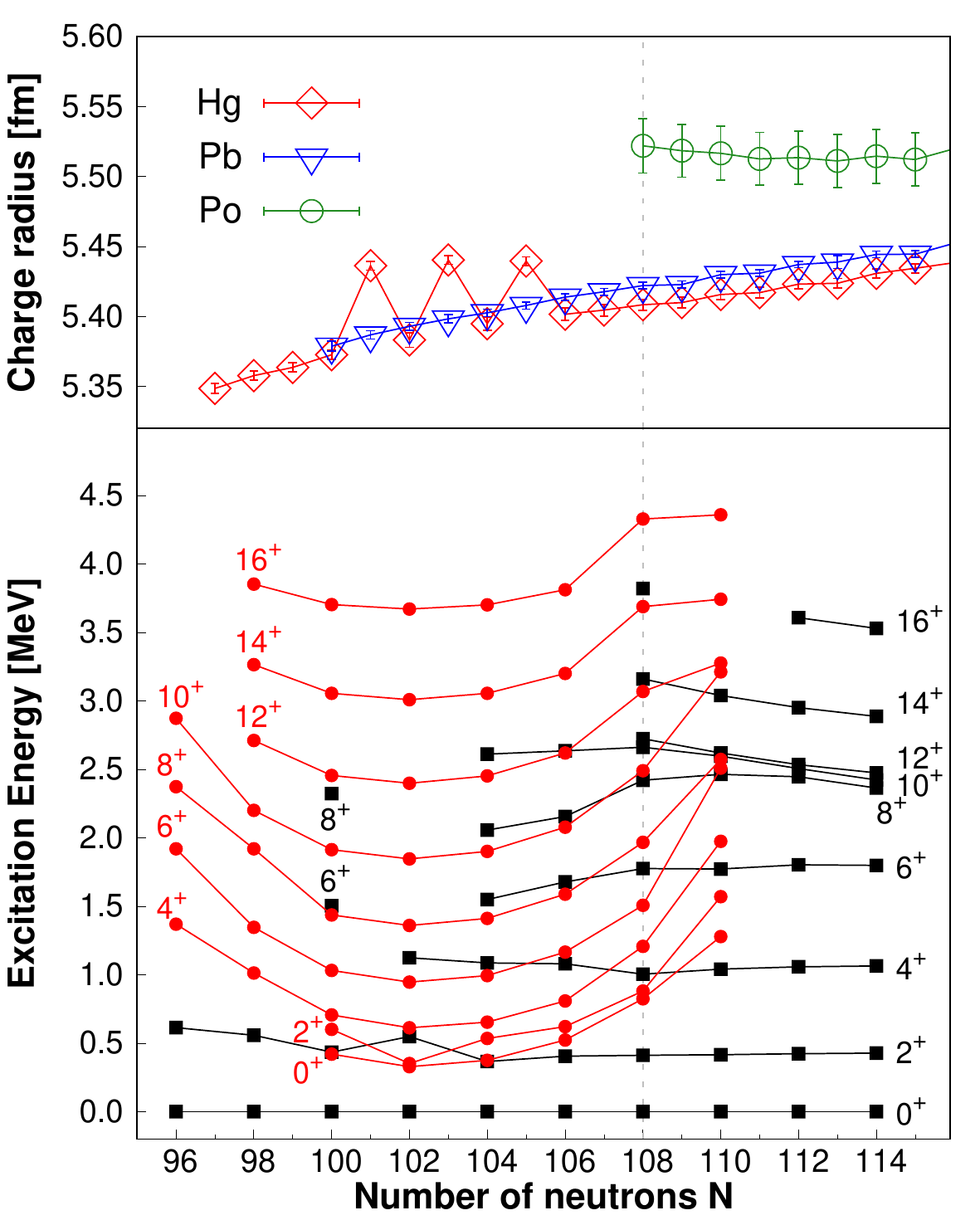}
\vspace{-6mm}
\caption{\label{fig:hints} (Color online) (top) Mean-square charge radius as a function of the neutron number for Po (green), Pb (blue) and Hg (red) isotopic chains. 
For the $Z=80$ isotopes, the large variation of the radius was attributed to the presence of different shapes. 
The values are taken from Refs.~\cite{bonn1971change, angeli2013, marsh2018characterization}. 
(bottom) Energy systematics of excited states in the neutron-deficient even-even mercury isotopes, showing (red circles) the assumed intruder and (black squares) ground-state bands. 
Data are taken from the National Nuclear Data Center database~\cite{nudat2}.}
\end{figure}

The first hint of shape coexistence in the Z$\approx$82 region came from studies of optical hyperfine structure in the neutron-deficient mercury isotopes~\cite{bonn1971change}. 
As shown in Figure~\ref{fig:hints} (top panel), significant staggering of the mean-square charge radius was observed for mass $181 \leq A \leq 185$, which was interpreted as resulting from the presence of two structures characterized by different deformation. 
This behavious of mean-square charge radii is unique in the nuclear chart. 
Recent laser-spectroscopy measurements~\cite{marsh2018characterization} demonstrated that the shape staggering is present down to $A=179$ ($N=99$), where the nucleus returns to sphericity in its ground state.

Another fingerprint of shape coexistence in the mercury isotopes is the observation of low-lying intruder bands built on second $0^+$ states, which are particularly close in energy to the ground state in the isotopes from $^{180}$Hg to $^{188}$Hg (see Figure~\ref{fig:hints} (bottom panel)).
From level energies within the bands, the deformations of these nuclei in the ground-state bands were estimated to be $\beta_2 \approx 0.1$, in contrast to $\beta_2 \approx 0.3$ obtained for the bands built on the $0^+_2$ states~\cite{cole1976isomerism, hamilton}. 
These structures tend to mix due to their proximity in energy.
The degree of their mixing was first estimated from the measured $\alpha$-decay hindrance factors, yielding a $3\%$ admixture of the deformed configuration in the ground state of $^{180}$Hg, while mixing of $16\%$ and $18\%$ was obtained for $^{182}$Hg and $^{184}$Hg, respectively~\cite{wauters1994alpha}.
These admixtures are consistent with those deduced from the $\rho^2(E0;0^+_2 \to 0^+_1)$ value for $^{188}$Hg and its upper limit for $^{186}$Hg~\cite{joshi1994lifetime}, and together they display a parabolic behavior as a function of neutron number with maximum mixing observed at the $N=104$ mid-shell, where the intruding structure comes the closest in energy to the ground state. 
The mixing of normal and intruder structures in Hg isotopes was also investigated by applying a phenomenological two-band mixing model to level energies in the observed rotational bands~\cite{dracoulis, gaffney2014shape}. 
The most recent study~\cite{gaffney2014shape} took into account the energies of newly identified non-yrast states in $^{180,182}$Hg and yielded a lower mixing between the $0^+$ states than that deduced from the $\alpha$-decay work of Ref.~\cite{wauters1994alpha}, but still with a maximum around $N=102-104$. 
The conclusions of this analysis~\cite{gaffney2014shape} for the $2^+$ states are, however, much different, suggesting an inversion of configurations of the $2^+_1$ state between $^{182}$Hg and $^{184}$Hg, with an almost maximum mixing strength ($51\%$) for the $2^+_1$ state in the mid-shell $^{184}$Hg nucleus. 
As for the 0$^+$ states, when moving away from $N=104$, both towards lighter and heavier nuclei, the configurations of the $2^+$ states become more pure. 
The importance of configuration mixing in the structure of $2^+$ states in neutron-deficient Hg nuclei is
further supported by enhanced $\rho^2(E0; 2^+_2 \to 2^+_1)$ transition strengths observed for $^{180}$Hg~\cite{page2011elspec}, $^{182}$Hg~\cite{rapisarda2017shape}, $^{184}$Hg~\cite{rapisarda2017shape} and $^{186}$Hg~\cite{cole1977shape, scheck2011elspec}.
Finally, the mixing of states with $J>2$ extracted from the perturbation of level energies in rotational bands in
$^{180-188}$Hg~\cite{gaffney2014shape} decreases with spin and it no longer displays a parabolic behavior as a function of neutron number, but rather a monotonic increase with mass (e.g. for the 4$^+$ states the
admixtures increase from $2\%$ in $^{180}$Hg to $20\%$ in $^{188}$Hg).
Currently, no E0 transition strengths are known in neutron-deficient Hg nuclei between the states of spin 4 and higher.

Further information on collective structures in neutron-deficient Hg nuclei was provided by measurements of $\gamma$-ray transition strengths. 
Lifetimes of excited states were measured for several neutron-deficient species, populated via fusion-evaporation reactions: the $^{178}$Hg~\cite{PhysRevC.99.054325}, $^{180,182}$Hg~\cite{grahn2009evolution, scheck2010lifetime} and $^{184,186}$Hg~\cite{rud1973lifetimes, proetel1974nuclear, gaffney2014shape} nuclei were studied using the Recoil-Distance Doppler-Shift (RDDS) method, while the Doppler-Shift Attenuation Method (DSAM) and the $\beta$-tagged fast-timing (FT) techniques were employed to investigate the $^{184}$Hg~\cite{ma1986structure} and $^{186,188}$Hg~\cite{joshi1994lifetime, olaizola2019ft} isotopes, respectively. 
Moreover, the lifetimes of low-lying states in $^{190-196}$Hg isotopes, which could not be obtained via RDDS method due to the presence of low-lying isomers,
were recently measured via the FT technique~\cite{esmaylzadeh2018lifetime, olaizola2019ft}. 
These studies were mostly limited to yrast states but, as the intruder band becomes yrast at low spin (see Figure~\ref{fig:hints} (bottom panel)), they yielded lifetimes of states in both coexisting bands in the even-mass $^{180-186}$Hg nuclei, providing strong support for the very different deformations of these two structures.

Finally, with the advent of radioactive ion-beam facilities, the even-mass $^{182-188}$Hg isotopes were investigated via low-energy Coulomb excitation, 
yielding magnitudes and signs of the reduced E2 matrix elements between the low-lying excited states~\cite{bree2014shape, wrzosek2019coulex}. 
Combined with the mixing coefficients of Ref.~\cite{gaffney2014shape}, these results provided a consistent picture of two distinct configurations (weakly-deformed oblate and strongly-deformed prolate) contributing in varying proportions to the observed low-lying states in $^{182-188}$Hg. 
Even though the simplest observables, such as the energy of the first-excited $2^+$ state and the $B(E2; 2^+_1 \to 0^+_1)$ value, are almost identical in $^{182-188}$Hg, the structures of $2^+_1$ states were demonstrated to be very different: the intruder configuration dominates in $^{182}$Hg, both configurations almost equally contribute in
$^{184}$Hg, and the normal configuration prevails for $^{186,188}$Hg. 
Unfortunately, due to strong correlations between the reduced matrix elements, the important role of E0 transitions between the $2^+$ states and the lack of sufficiently precise lifetimes, branching and mixing ratios for the nuclei of interest, it was not possible to determine any spectroscopic quadrupole moments from this study, except for the $Q_s(2^+_1) = 0.8^{+0.5}_{-0.3}$ eb in $^{188}$Hg,
consistent with an oblate deformation of this state. 


The first interpretation of the intruder band was given in the work of Praharaj and Khadkikar~\cite{praharaj1980shape}, who 
performed Hartree-Fock calculations for even-mass mercury isotopes from $A=184$ to $A=204$ and obtained $\beta_2$ deformation parameters for the two coexisting structures. 
In their calculations, all these nuclei present oblate ground-state bands, with a maximum of the deformation for $A=186$ $(\beta_2 = 0.117)$, while the excited  bands are prolate-deformed. 
In particular, small mixing was predicted between the two bands in $^{188}$Hg.
More recently, Nik\v{s}i\'{c} and collaborators performed relativistic Hartree-Bogoliubov (RHB) calculations~\cite{niksic2002} and predicted the ground-state band of Hg isotopes to be weakly-deformed oblate, due to the two-proton hole in the $Z=82$ shell. 
For the isotopes close to the neutron mid-shell, including $^{188}$Hg, this oblate ground-state band was predicted to be crossed by
an intruding prolate-deformed band, related to $4p-6h$ proton excitations into the $h_{9/2}$ and $f_{7/2}$ orbitals, and the two structures were expected to be strongly mixed.
Similar conclusions could also be reached from beyond-mean-field (BMF) and interacting-boson model (IBM) calculations that were summarized in the works of Bree et al.~\cite{bree2014shape} and Wrzosek-Lipska et al.~\cite{wrzosek2019coulex}. 
In these works, the BMF approach predicts for $N \ge 106$ a weakly-deformed ground-state band coexisting with an excited prolate band characterized by a stronger deformation. 
For nuclei with $100 \le N \le 104$ the two bands cross and the ground state is expected to be predominantly prolate, while the first excited $0^+$ state is predicted to have equal contributions of the oblate and the prolate configuration. 

The intruder structure of $^{188}$Hg appears at excitation energy slightly higher than in $^{182-186}$Hg, but it also becomes yrast very quickly, starting from spin $6^+$. 
The degree of mixing between the two coexisting structures is expected to be lower than at mid-shell, but the detailed predictions of the models significantly differ. 
As $^{188}$Hg is less exotic and thus more accessible for high-precision spectroscopy than the Hg isotopes in the nearest vicinity of $N=104$, the paucity of information about its electromagnetic structure, in particular that of higher-spin states, is surprising. 
The transition probabilities in $^{188}$Hg were studied via lifetime measurements using the FT technique~\cite{joshi1994lifetime, olaizola2019ft} and via Coulomb excitation~\cite{bree2014shape, wrzosek2019coulex}.
However, the existing information is mostly limited to yrast states and the results on the $2_1^+$ lifetime are not consistent. 
In order to obtain a clear picture of shape coexistence in $^{188}$Hg and a deeper insight into configuration mixing in Hg nuclei, precise determination of the reduced transition probabilities between the low-lying states, belonging to both structures, is mandatory. 

In the present paper, the lifetime measurement of the excited states in both the ground-state and intruder bands
in $^{188}$Hg is presented.
The results are discussed and interpreted 
in the context of new beyond-mean-field calculations performed via the symmetry-conserving configuration-mixing and the five-dimensional collective Hamiltonian methods. 

\begin{figure*}[t]
\includegraphics[width=0.9\textwidth]{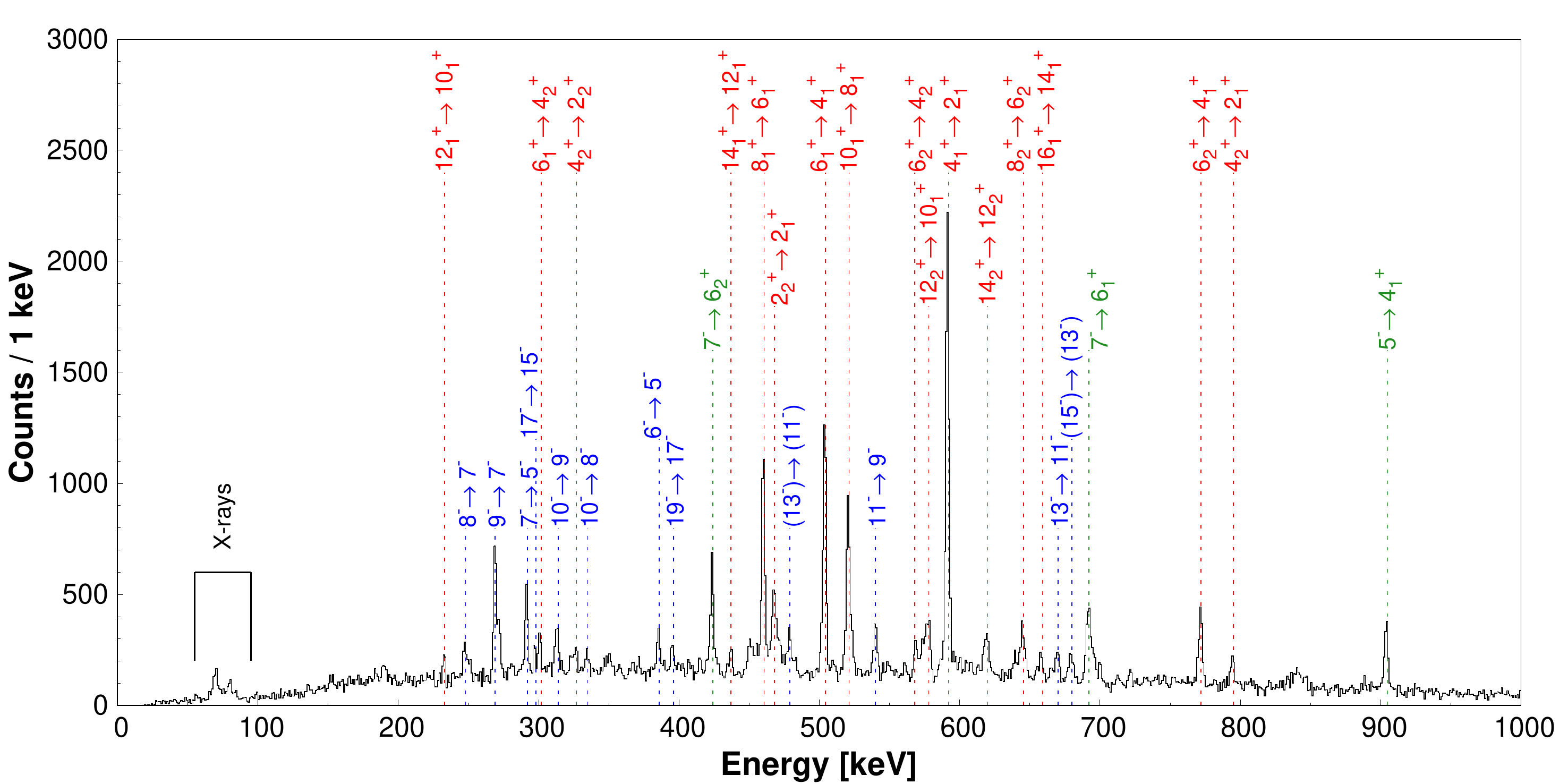}
\includegraphics[width=0.8\textwidth]{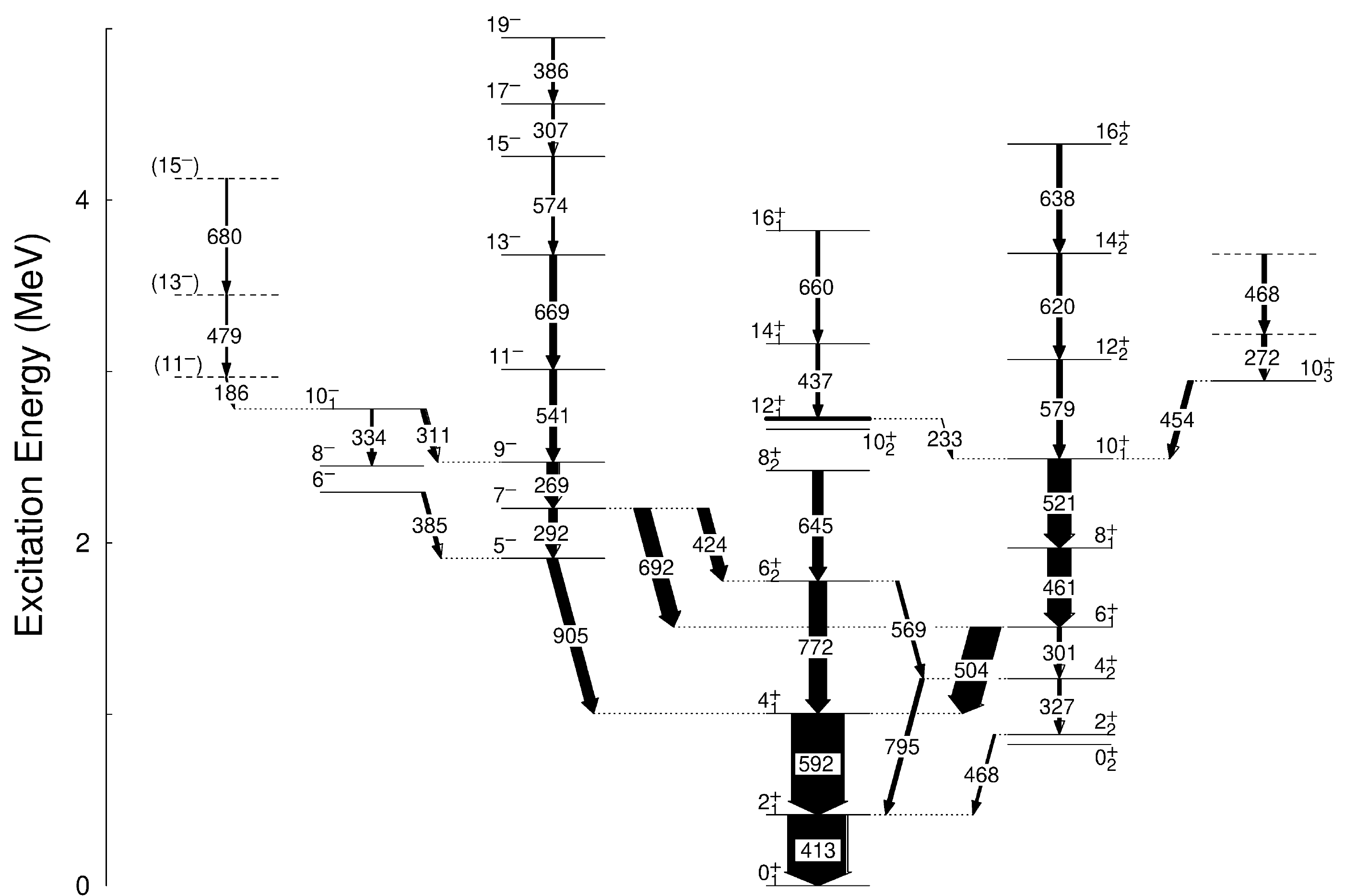}
\vspace{-1mm}
\caption{\label{fig:188Hg} (Color online) (top) Background-subtracted $\gamma$-ray energy spectrum from the GALILEO detectors at $90^\circ$, obtained by gating on the 413-keV $2_1^+ \to 0_{g.s.}^+$ transition and requiring the coincidence with at least one neutron. 
The statistics is the sum of the \textit{Exp.1} and \textit{Exp.2} datasets. 
The identified transitions in the $^{188}$Hg positive-parity bands are marked in red, those in and between negative-parity bands are highlighted in blue, while those between positive-parity and negative-parity bands are in green. 
(bottom) Partial level scheme of $^{188}$Hg, reporting the $\gamma$-ray transitions observed in the present measurement. 
The arrow widths represent the efficiency-corrected transition yields.}
\end{figure*}

\section{Experimental details}

Two experiments were performed, using different fusion-evaporation reactions, to populate the excited states in $^{188}$Hg. 
In the first one, a $^{34}$S beam at 185-MeV energy impinged on a 600 $\mu$g/cm$^2$ thick target of $^{160}$Gd, evaporated onto a 2.5 mg/cm$^2$ thick $^{181}$Ta foil. 
The second one used a $^{34}$S beam at the energy of 165 MeV and a 600 $\mu$g/cm$^2$ thick target of $^{158}$Gd
evaporated on an identical foil. 
In both experiments, the targets were oriented such that the beams first passed through the Ta foils before striking the target material.
In the following, these two measurements will be referred to as \textit{Exp.1} and \textit{Exp.2}, respectively. 
As the cross sections for individual reaction channels were different for \textit{Exp.1} and \textit{Exp.2}, this approach provided a better control of possible contamination of the $^{188}$Hg data by other reaction channels. 
The $^{34}$S beam was provided by the Tandem-ALPI accelerator complex~\cite{signorini1984acceptance, danielli1996} of the Laboratori Nazionali di Legnaro (Italy). 
The $\gamma$ rays resulting from the reactions were detected by the GALILEO spectrometer, an array of 25 Compton-shielded HPGe detectors, arranged into 3 rings at backward angles (152$^{\circ}$, 129$^{\circ}$, 119$^{\circ}$) and one ring at 90$^{\circ}$~\cite{valiente2014GALILEO} with respect to the beam direction. 
The neutrons evaporated in the reaction were detected using the Neutron Wall array~\cite{skeppstedt1999}, composed of 45 liquid scintillators placed at forward angles with respect to the beam direction.
The use of Neutron Wall was necessary to discriminate between the events of interest, expected in coincidence with at least one neutron, and the Coulomb-excitation background~\cite{zanon2017shape}. 
Figure~\ref{fig:188Hg} (top panel) shows the $\gamma$-ray energy spectrum of $^{188}$Hg, obtained in coincidence with at least one neutron and gated on the $2_1^+ \to 0_{g.s.}^+$ transition in $^{188}$Hg.
The partial level scheme, showing all $\gamma$-ray transitions observed in the present study, is presented in the bottom panel of Figure~\ref{fig:188Hg}. 
More details about the pre-sorting of the data can be found in Ref.~\cite{zanon2019shape}.

For the lifetime measurements, the RDDS technique~\cite{dewald2012developing} was used by employing the GALILEO plunger~\cite{MULLERGATERMANN201995} with a 11 mg/cm$^2$ thick $^{197}$Au stopper mounted after the target. 
For each $\gamma$-ray transition two components were observed, related to the radiation emitted before and
in the $^{197}$Au foil: the $\gamma$ rays emitted in-flight after the target are Doppler shifted, while those emitted after the implantation in the stopper are detected at the proper energy. 
From the energy difference between the in-flight and the stopped components of the $\gamma$-ray transitions, the average velocity of the $^{188}$Hg evaporation residue (ER) was determined, being $\beta = 1.71(8)\%$ for \textit{Exp.1} and $\beta = 1.59(1)\%$ in \textit{Exp.2}. 
Considering the velocity of the nucleus of interest, seven target-stopper distances in the range 20-600 $\mu$m were used during \textit{Exp.1} and seven in the range 7-2000 $\mu$m during \textit{Exp.2}, to measure lifetimes between few and tens of picoseconds.
Specifically, in the first experiment the distances were optimized to get information mostly on the $2_1^+$ excited state, while for the second they were selected to extend the measurement also to shorter-lived states.

\section{Lifetime analysis}

\begin{figure}[b]
\includegraphics[width=0.48\textwidth]{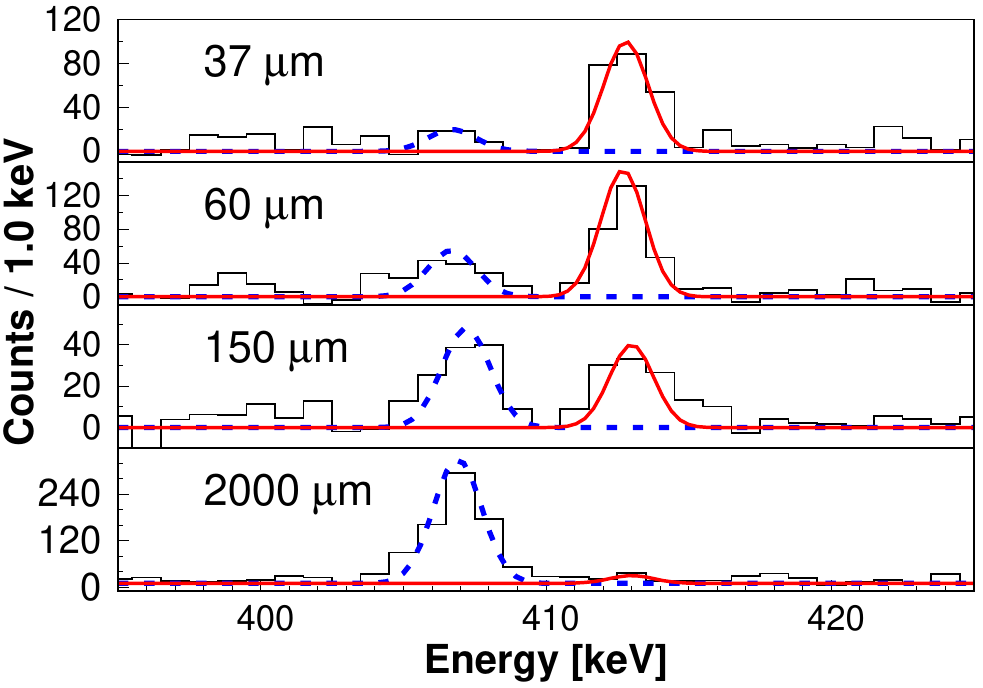}
\vspace{-6mm}
\caption{\label{fig:spec_2p0p} (Color online) Background-subtracted $\gamma$-ray energy spectra of $^{188}$Hg for different target-to-stopper distances for the \textit{Exp.2} dataset and the GALILEO detectors at 152$^\circ$. 
The spectra are expanded in the vicinity of the $2_1^+ \to 0_1^+$ transition and obtained by gating on the in-flight component of the $4_1^+ \to 2_1^+$ transition and requiring the coincidence with at least one neutron. 
The in-flight and stopped components are indicated by dashed-blue and solid-red lines, respectively.}
\end{figure}

In order to avoid the effects of unobserved feeding transitions and to reduce the possibility of contamination from different reaction channels, the lifetime measurements were performed using the $\gamma$-$\gamma$ coincidence procedure, gating on the in-flight component of the most intense feeding transition. 
Moreover, in order to discriminate against the events resulting from the Coulomb excitation of both $^{197}$Au stopper and $^{181}$Ta target fronting, all analyzed $\gamma$-$\gamma$ matrices were constructed by requiring the coincidence with at least one neutron identified in the Neutron Wall array. 
Figure~\ref{fig:spec_2p0p} shows the evolution of the intensities of the in-flight and stopped peaks of the $2_1^+ \to 0_{g.s.}^+$ transition as a function of the target-degrader distance, after gating on the $4_1^+ \to 2_1^+$ in-flight component. 
The lifetimes of the states were extracted using the NAPATAU software~\cite{napatau}, applying the Differential Decay Curve Method (DDCM)~\cite{dewald2012developing} by fitting the area of both the in-flight ($I^{if}_i$) and the stopped ($I^{st}_i$) components with a polynomial piecewise function. 
These intensities were scaled according to an external normalization, given by the area of the 136-keV $\gamma$-ray peak of $^{181}$Ta, coming from the Coulomb excitation of the target fronting.
This choice for the normalization was due to the fact that the number of counts in this peak is not only proportional to the beam intensity and duration of the run, but it also provides a measure of possible degradation of the target during the experiment.

The lifetime $\tau_i$ should be the same for each $i$-th target-stopper distance and it is obtained as 

\begin{equation}
\tau_i = \frac{I^{st}_i -\Sigma_j \left( Br {\,} \alpha {\,} I^{st}_i \right)_j}{\frac{d}{dt}I^{if}_i} \, ,
\end{equation}
where the summation is extended over the $j$ feeding transitions, each with a certain branching ratio ($Br$) and parameter $\alpha$, which includes the efficiency correction and the angular correlation between the transition of interest and the feeding one. 
In the case of the $\gamma$-$\gamma$ coincidence procedure with gating on the in-flight component of the feeding transition, the contributions from feeding transitions are eliminated and this term is null.
The final result is given by the weighted average of the lifetimes within the sensitive region of the technique, i.e. where the derivative of the fitting function is largest. 

\begin{figure*}[t]
\includegraphics[width=\textwidth]{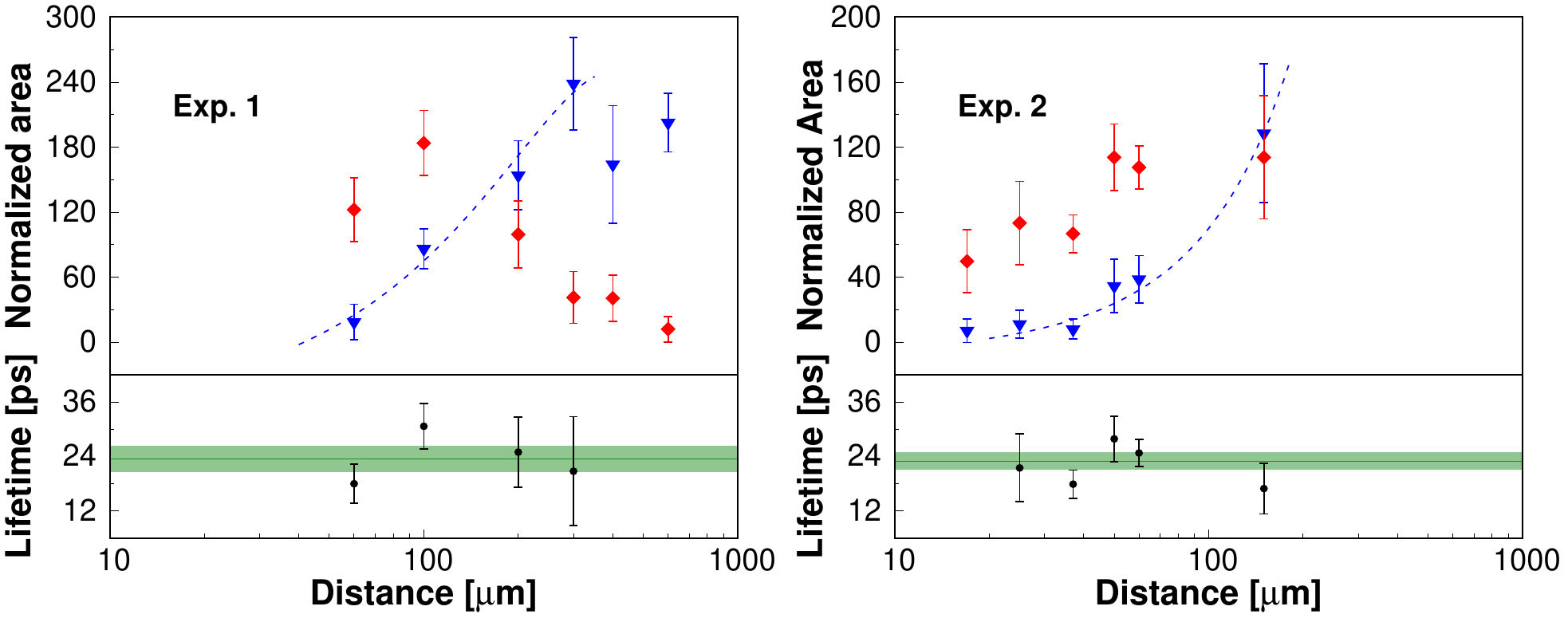}
\vspace{-6mm}
\caption{\label{fig:2p0p} (Color online) DDCM analysis for the lifetime measurement of the $2_1^+$ excited state, after gating on the in-flight component of the $4_1^+ \rightarrow 2_1^+$ transition. 
(top) Area of the in-flight (blue) and stopped (red) components, normalized to the area of the 136-keV $\gamma$-ray peak of $^{181}$Ta. 
The dashed lines represent a fit to the in-flight-component points in the sensitive region of the technique. (bottom) Corresponding lifetimes obtained for individual distances. 
The solid line denotes the weighted average of the lifetimes, while the filled area corresponds to 1$\sigma$ uncertainty.}
\end{figure*}

The lifetimes of the $J^\pi = 2_1^+, \, 4^+_1, \, 6_1^+, \, 8_1^+, \, 10_1^+, \, 14_1^+$ and $16_1^+$ excited states in $^{188}$Hg were extracted via the DDCM by gating on the in-flight component of the $(J +2)^\pi \to J^\pi$ feeding transition. 
The lifetimes of these states were obtained using data for each of the three GALILEO rings separately and then the weighted averages were calculated, except for the $16^+$ state where in order to obtain sufficient  statistics data from the three rings had to be summed. 
In Figure~\ref{fig:2p0p} the DDCM analysis performed for the $2^+_1$ state is presented for \textit{Exp.1} and \textit{Exp.2}, showing the results for the detectors at 129$^\circ$ and 152$^\circ$, respectively. 
The lifetimes obtained from the two experiments and data from the three HPGe rings are in a good agreement. 
Thus, the weighted averages of the results obtained for the different detector angles were calculated, leading to $\tau(2_1^+) = 25(2)$ ps for \textit{Exp.1}, $\tau(2_1^+) = 24(1)$ ps for \textit{Exp.2}. 
The lifetime of the $4^+_1$ state could be determined only from the \textit{Exp.2} dataset, which yielded $\tau(4^+_1) = 1.9(8)$ ps.
These results are in agreement with those given in the literature~\cite{bree2014shape, wrzosek2019coulex, olaizola2019ft}. 

The stopped components of the $\gamma$-ray decays from the $12_2^+$ state were observed only at the two shortest distances employed in \textit{Exp.2} only, implying a very short lifetime for the $12_2^+$ state. 
Using these two data points, a lifetime of $\approx 2$ ps was extracted, which would yield a transitional quadrupole moment consistent with the trend observed for the high-spin members of the intruder band (see
discussion in Sec.~\ref{sec.discussion}).
However, due to an insufficient number of experimental points in the sensitive range of the method, we adopt a more conservative upper limit of 4 ps for this state. 

\begin{figure}[b]
\includegraphics[width=0.48\textwidth]{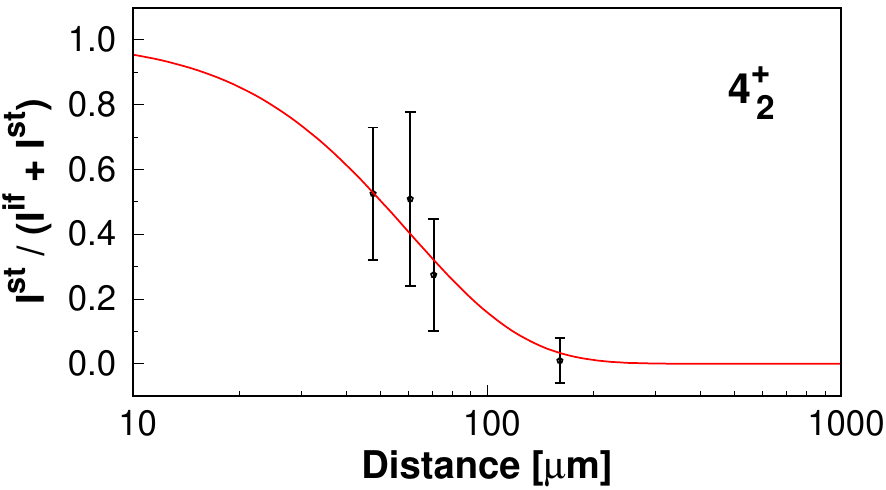}
\vspace{-5mm}
\caption{\label{fig:4p} (Color online) Decay curve of the $4_2^+$ excited state in $^{188}$Hg as a function of the target-stopper distance, obtained for the \textit{Exp.2} dataset by gating on the in-flight component of the $8_1^+ \to 6_1^+$ transition (see text). 
The red line represents the fitted decay curve, assuming  $\tau(6_1^+) = 5.1(5)$ ps.}
\end{figure}

The presence of the 154-ns $12_1^+$ isomeric state prevented the investigation of the $J^\pi=10_2^+, 8_2^+, 6_2^+$ states with the RDDS technique.
From the \textit{Exp.1} and \textit{Exp.2} measurements, an upper limit of 10 ps can be set for the $6_2^+$ excited state, since in the spectra gated on both the 645-keV $8_2^+ \to 6_2^+$ and the 424-keV $7^- \to 6_2^+$
transitions only the in-flight component of the $6_2^+ \to 4_1^+$ transition was observed clearly for the data obtained for the longer plunger distances.

Due to the insufficient level of statistics for the in-flight component of the 301-keV $6^+_1 \rightarrow 4^+_2$ transition, it was not possible to extract the lifetime of the $4^+_2$ state via the DDCM. 
Thus, for the \textit{Exp.2} dataset, a gate on the 460-keV $8^+_1 \rightarrow 6^+_1$ transition was set and the lifetime was extracted via the Decay-Curve Method (DCM), using a second-order Bateman equation and taking into account the lifetime of the $6_1^+$ state. 
Figure~\ref{fig:4p} shows the decay curve for the $4_2^+$ state, resulting in a lifetime of $\tau(4^+_2) = 8(3)$ ps. 

\begin{table}[h]
\caption{\label{tab:Results} Lifetimes (in picoseconds) of the $J_i^\pi$ excited states in $^{188}$Hg. 
The weighted averages of the results from the three GALILEO rings are given and compared with the literature values~\cite{bree2014shape, wrzosek2019coulex, olaizola2019ft}. 
The reported uncertainties are statistical only.}
\begin{ruledtabular}
\begin{tabular}{c c c c}
$J_i^\pi$	&	\textit{Exp.1} &	\textit{Exp.2}	&	\textit{Previous}\\
\hline
$2^+_1$	&	25(2)	&	24(1)	&	19(3)\\
$4^+_1$	&	-		&	1.9(8)	&	2.3(2)\\
$4^+_2$	&	$<10$	&	8(3)	&	$<58$\\
$6_1^+$	&	5.5(6)	&	5.1(5)	&	$<14$	\\
$6_2^+$	&	$<10$	&	$<10$	&	$<30$ \\
$8_1^+$	&	-		&	2.7(2)	&	-\\
$10_1^+$&	-		&	2.1(2)	&	-\\
$12_2^+$&	-		&	$<4$	&	-\\
$14_1^+$&	18(4)	&	19(3)	&	-\\
$16_1^+$&	$<20$	&	12(3)	&	-\\
\end{tabular}
\end{ruledtabular}
\end{table}

Table~\ref{tab:Results} summarizes the experimental results from the two datasets. 
The reported uncertainties represent the $1\sigma$ statistical error, given by the adopted procedures and the weighted average of the results from the individual GALILEO rings.
Additionally, a systematic uncertainty (typically $\leq 3 \%$), accounting for the choice of the fitting function and relativistic effects~\cite{dewald2012developing}, should be considered.
Because of the lack of shorter distances investigated in \textit{Exp.1}, it was possible to study the lifetimes of the $2_1^+$, $6_1^+$ and $14_1^+$ excited states only. 
Similar to the $6_2^+$ case discussed above, only an upper limit can be set for the $4_2^+$ and $16_1^+$ states.

\section{Discussion}
\label{sec.discussion}

From the weighted average of lifetimes ($\bar{\tau}$) measured in \textit{Exp.1} and \textit{Exp.2}, the corresponding $B(E2)$ values can be extracted, which can be related to the deformation parameters using simple rotational approximations.
In the axially symmetric rotor model, the transition quadrupole moment $Q_t$ is related to the E2 transition probability via

\begin{equation}
B(E2; J_i \to J_f) = \frac{5}{16 \pi} \langle J_i, K, 2, 0 | J_f, K \rangle^2 Q_t^2 ~,
\label{eq:be2_1}
\end{equation}
where $J_f = J_i -2$ and $\langle J_i, K, 2, 0 | J_f, K \rangle$ is the Clebsch-Gordan coefficient. 
$K$ is the spin projection along the symmetry axis of the nucleus, conserved in the case of axial symmetry (thus it is equal to 0 for the considered states in both coexisting structures in $^{188}$Hg). 
Under this assumption, $Q_t$ does not change within a band and it is equal to the intrinsic quadrupole moment $Q_0$, which is related to the $\beta_2$ deformation parameter via 

\begin{equation}
Q_0 = \frac{3}{\sqrt{5 \pi}} Z R_0^2 \beta_2 (1 +0.36 \beta_2) ~,
\label{eq:q0}
\end{equation}
where $R_0 = 1.20 A^{1/3}$. 

\begin{figure}[h]
\includegraphics[width=0.48\textwidth]{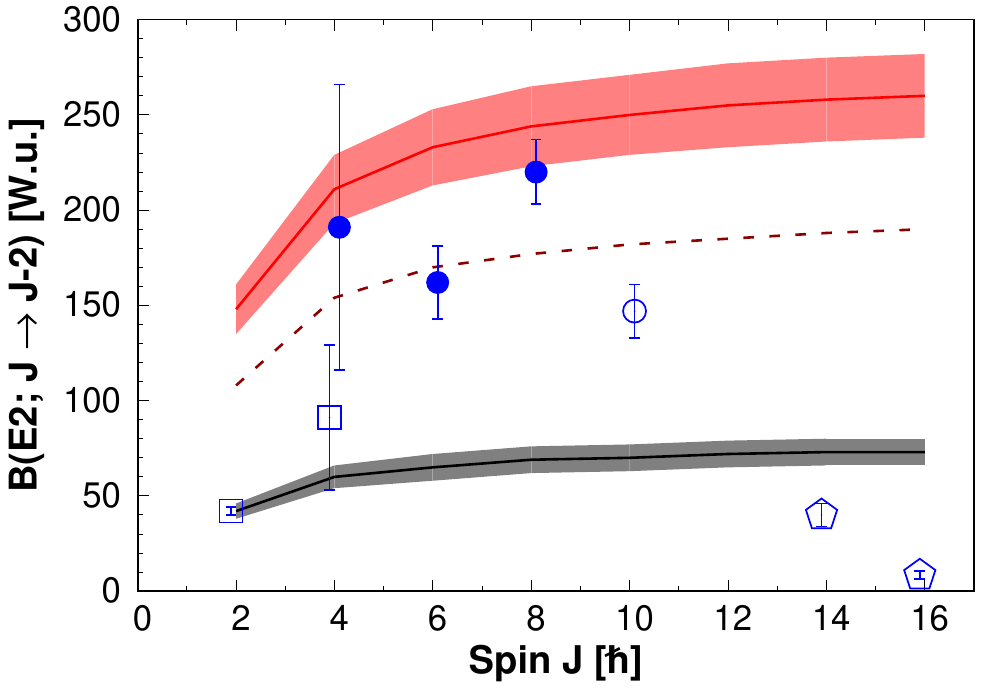}
\vspace{-5mm}
\caption{\label{fig:B(E2)_mixing} (Color online) Measured B(E2) values as a function of spin are compared with those calculated from the two-band mixing model~\cite{LANE1995129}. 
The black (red) solid line denotes pure normal (intruder) structures transition strengths, while the related shaded areas correspond to their error bars. 
The B(E2) values for the pure intruder configuration, estimated from the axially symmetric rotor model (see text) are presented with a dashed red line. 
The experimental values are presented with open squares for the ground-state band, with solid circles for the $J<10$ states of the intruder band, with an open circle for the $J=10$ state of the intruder band (see text) and with open pentagons for the almost spherical band above the $12_1^+$ isomer.}
\end{figure}

In Table~\ref{tab:B(E2)} the $B(E2)$ values and the corresponding $Q_t$ and $\beta_2$ values are reported for the in-band transitions. 
Based on the excitation energies of the states and the $Q_t$ values, three different configurations can be identified: a slightly-deformed ground-state structure below the $12_1^+$ isomeric state, an almost spherical structure above it and finally the deformed intruder band. 
For the latter, within the experimental uncertainties, the quadrupole moments tend to be constant for the $J \ge 4$ states ($Q_t \approx 5.88$ eb, which corresponds to $\beta_2 \approx 0.21$). 
Due to its low magnitude and slow evolution as a function of spin~\cite{gaffney2014shape}, the mixing between the ground-state and intruder bands only weakly affects the intra-band transition probabilities, leading to a rather constant trend of the quadrupole moments.
Therefore a first estimate of reduced transition probabilities in the pure intruder structure can be obtained using the average $Q_t$ value for higher-spin states in the intruder band, as shown in Figure~\ref{fig:B(E2)_mixing}.

Since there is substantial experimental evidence for the mixing between the ground-state and the intruder bands (e.g. significant $\rho^2(E0)$ values, intense inter-bands transitions), it is reasonable to assume that the two structures are not axially symmetric. 
Thus, assuming that the $J \geq 4$ intruder states result from a rotation of a non-axial nucleus characterized by two deformation parameters $(\beta_2 , \gamma)$, one can relate the measured B(E2) values to the intrinsic quadrupole moment $Q_0$ using the following extension of Equation \eqref{eq:be2_1}, as proposed for example in Ref.~\cite{petkov1998ba128}: 

\begin{equation}
\begin{split}
&B(E2;J \to J-2) = \frac{5}{8\pi} Q_0^2 \frac{J(J-1)}{(2J-1)(2J+1)} \\
&\times \left[ cos(\gamma+30^\circ) -\frac{\left\langle K^2 \right\rangle}{(J-1)J}cos(\gamma-30^\circ) \right]^2 \, .
\end{split}
\label{be2_2}
\end{equation}
Applying the described model to the measured B(E2) values for $J \geq 4$, one obtains for the pure intruder configuration $Q_0 = 5.90(22)$ eb (corresponding to $\beta_2=0.21(1)$) and $\gamma=14(2)^\circ$. 
Here, we again assumed $K=0$ for all states in the intruder band, as the departure from axial symmetry is not large and $K$ is likely to be approximately a good quantum number. 

In order to shed light on the nature of the coexisting structures, accounting for the mixing between them, the experimental reduced transition probabilities are further compared with a band-mixing model~\cite{LANE1995129} and with beyond-mean-field calculations.

\begin{table*}[t]
\caption{\label{tab:B(E2)} Comparison between the reduced transition probabilities $B(E2; J_i^\pi \to J_f^\pi)$, extracted from the measured lifetimes, and the theoretical values predicted by 5DCH and SCCM for $^{188}$Hg. 
The transition quadrupole moment $Q_t$ and the $\beta_2$ deformation parameter are extracted assuming an axially symmetric rotor model. 
The experimental values marked with $^{(a)}$ and the branching ratios ($Br$) are taken from Ref.~\cite{KONDEV20181}.}
\begin{ruledtabular}
\begin{tabular}{c c c c c c c c c}
\multirow{2}{*}{$J_i^\pi \rightarrow J_f^\pi$}	&	\multirow{2}{*}{$E_{\gamma}$ [keV]}	&	\multirow{2}{*}{$\bar{\tau}$ [ps]}	&	\multirow{2}{*}{$Br$}	&	\multicolumn{3}{c}{$B(E2)$ [W.u.]} & \multirow{2}{*}{$|Q_t|$ [eb]} & \multirow{2}{*}{$\beta_2$}\\ 
\cline{5-7}
		&	&	&	&	\textit{Exp.} & \textit{5DCH} & \textit{SCCM}	&	&	\\
\hline
$2^+_1 \rightarrow 0^+_1$ 	& 	$413$	&	24.4(9)					&	1.00	&	44(2)	&	82	&	60		&	3.66(10)	&	0.128(4)\\
$4^+_1 \rightarrow 2^+_1$ 	&	$592$	&	1.9(8)					&	1.00	&	92(38)	&	109	&	94		&	4.5(13)		&	0.158(47)\\
$6^+_2 \rightarrow 4^+_1$ 	&	$772$	&	$<10$					&	0.78(1)	&	$>4$	&	22	&	113		&	$>0.85$		&	$>0.03$\\
$12_1^+ \rightarrow 10^+_2$	&	$62$	&	$^{(a)}$222(29) 10$^3$	&	1.00	&	1.3(2)	&	-	&	-		&	0.50(5)		&	0.017(2)\\
$14_1^+ \rightarrow 12^+_1$	&	$437$	&	19(3)					&	1.00	&	40(6)	&	-	&	-		&	2.73(31)	&	0.095(11)\\
$16_1^+ \rightarrow 14^+_1$	&	$659$	&	12(3)					&	1.00	&	8(2)	&	-	&	-		&	1.24(22)	&	0.043(8)\\
$4^+_2 \rightarrow 2^+_2$ 	&	$327$	&	8(3)					&	0.46(2)	&	191(76)	&	215	&	316		&	6.56(184)	&	0.23(6)\\
$6_1^+ \rightarrow 4^+_2$ 	&	$301$	&	5.3(3)					&	0.16(1)	&	156(14)	&	295	&	355		&	5.65(36)	&	0.20(2)\\
$8_1^+ \rightarrow 6^+_1$ 	&	$461$	&	2.7(2)					&	1.00	&	220(16)	&	347	&	382		&	6.56(35)	&	0.23(1)\\
$10_1^+ \rightarrow 8^+_1$ 	&	$521$	&	2.1(2)					&	1.00	&	155(15)	&	330	&	-		&	5.42(37)	&	0.19(1)\\
$12_2^+ \rightarrow 10^+_1$	&	$578$	&	$<4$					&	1.00	&	$>48$	&	-	&	-		&	$>3.00$		&	$>0.10$\\
$2^+_2 \rightarrow 0^+_1$ 	& 	$881$	&	$^{(a)}$203(45)			&	0.60(3)	&	0.08(2)	&	0.20	&	0.4		&	-	&	-\\
$6_1^+ \rightarrow 4^+_1$ 	&	$504$	&	5.3(3)					&	0.79(1)	&	57(4)	&	13	&	1.5		&	-	&	-\\
\end{tabular}
\end{ruledtabular}
\end{table*}

\subsection{Two-band mixing}

The assumption of a purely rotational character of the high-spin intruder states is in line with the work of Gaffney et al.~\cite{gaffney2014shape} where, considering a spin-independent interaction between two rotational structures and employing the method of Ref.~\cite{LANE1995129}, the mixing amplitude ($\alpha_{mix}$) for the excited states up to spin $10 \hbar$ were extracted from the excitation energies.
The $6_1^+$, $8_1^+$ and $10_1^+$ excited states were estimated to have admixture of the normal structure at the level of $6.1\%$, $1.2\%$ and $0.5\%$, respectively. 

By applying the two-band mixing model~\cite{LANE1995129} to the transition probabilities determined from the presently measured lifetimes, information on the pure normal and intruder configurations can be obtained. 
Indeed, such a two-state mixing model is a simple approach to interpret the properties of physical states based on the mixing of different intrinsic configurations. 
As extensively discussed in the work of Cl\'{e}ment et al.~\cite{clement2007kr}, 
the experimentally observed states $| J_{1,2}^\pi \rangle$ can be written as a linear combination of the intrinsic pure prolate ($| J_{pr}^\pi \rangle$) and oblate states ($|J_{ob}^\pi \rangle$):

\begin{equation}
\begin{split}
| J_1^\pi \rangle &= \alpha_{mix} | J_{ob}^\pi \rangle + \sqrt{1-\alpha^2_{mix}} | J_{pr}^\pi \rangle\\
| J_2^\pi \rangle &= -\sqrt{1-\alpha^2_{mix}} | J_{ob}^\pi \rangle + \alpha_{mix} | J_{pr}^\pi \rangle 
\end{split}
\, ,
\label{eq-mixing}
\end{equation}

\noindent
where the pure structures are assumed to be orthogonal but their mixing gives rise to the observed inter-band transitions. 

Based on the excitation energy of the states belonging to the ground-state and intruder bands (see Figure~\ref{fig:188Hg}), it seems that only the $J \leq 8$ levels should be considered when applying the two-band mixing model. 
Indeed, in the ground-state band the rotational-like pattern is maintained up to the $8_2^+$ state, while the $10_2^+$ state has an energy so close to that of the $12_2^+$ isomer that it seems to belong to a different configuration. 
In the intruder band, instead, the $10_1^+$ state is fed by a 456-keV transition from another $10^+$ level, so one cannot exclude the possibility of it being mixed with this third structure. 
Thus, for the $J \leq 8$ states one can estimate the deformation of the unperturbed configurations from the experimental $B(E2; J \to J-2)$ values, adopting the experimentally-deduced mixing strengths of Ref.~\cite{gaffney2014shape}.
The resulting normal structure is characterized by $|Q_0^n| = 3.66(37)$ eb ($\beta_2 \approx 0.13$) and the intruder one by $|Q_0^i| = 6.90(57)$ eb ($\beta_2 \approx 0.25$). 
The reduced transition probabilities, corresponding to the unperturbed structures, are compared with the experimental results in Figure~\ref{fig:B(E2)_mixing}: while for the $4_2^+ \to 2_2^+$ transition the large uncertainty does not allow to make any conclusion on the mixing, the strengths of the $2_1^+ \to 0_{g.s.}^+$, $4_1^+ \to 2_1^+$, $6_1^+ \to 4_2^+$ and $8_1^+ \to 6_1^+$ transitions confirm the estimated $\alpha_{mix}$ of Ref.~\cite{gaffney2014shape}; on the other hand, the disagreement between the $B(E2; 10_1^+ \to 8^+_1)$ value and the prediction for the unperturbed intruder structure supports the hypothesis previously discussed, suggesting a more complex nature of the $10^+$ states that goes beyond the two-band mixing model.

\subsection{Beyond-mean-field calculations}

\begin{figure*}[t]
\centering
\includegraphics[width=\textwidth]{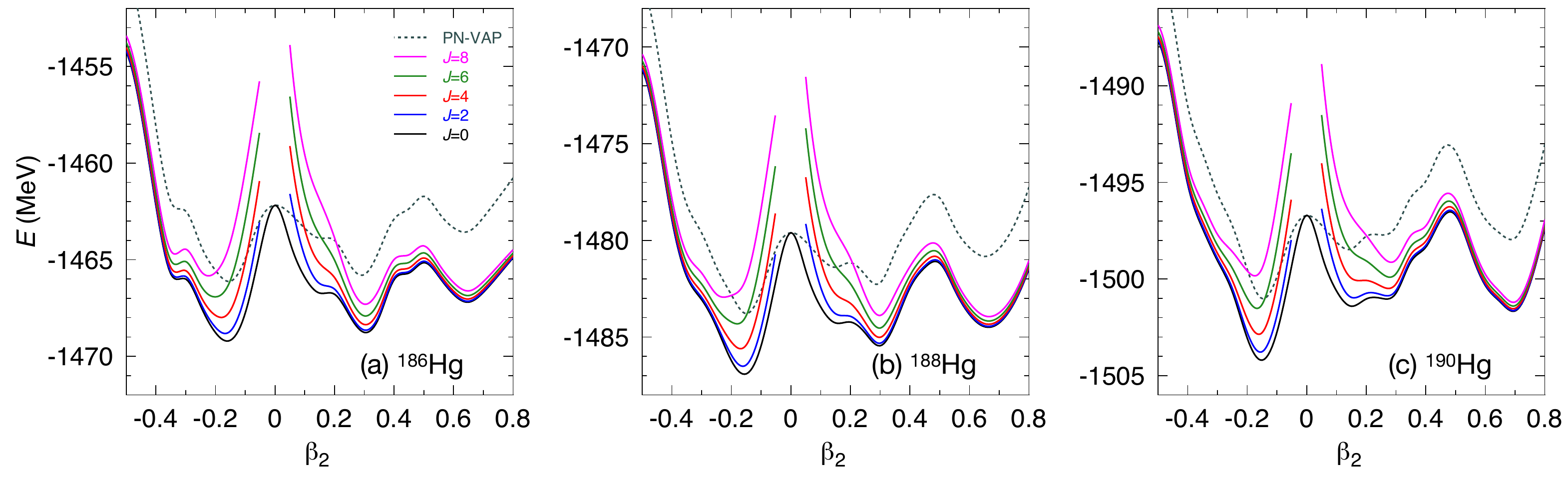}
\vspace{-7mm}
\caption{\label{fig:SCCM_PES} (Color online) PN-VAP (dashed lines) and PNAMP (continuous lines) energy curves as a function of the axial quadrupole deformation $\beta_{2}$ for the (a) $^{186}$Hg, (b) $^{188}$Hg and (c) $^{190}$Hg isotopes. 
The results are obtained with the Gogny-D1S interaction within the SCCM approach.}
\end{figure*}
\begin{figure*}[t]
\centering
\includegraphics[width=\textwidth]{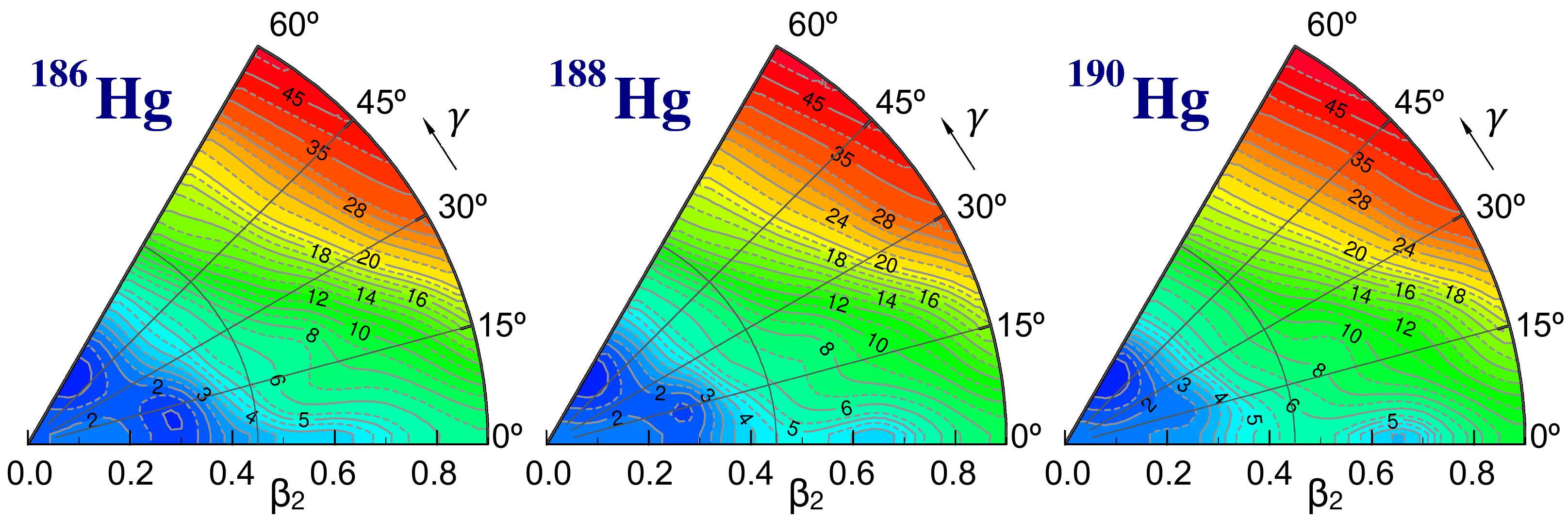}
\vspace{-7mm}
\caption{\label{fig:5DCH_PES} (Color online) Potential energy surfaces as a function of the $\beta_2$ and $\gamma$ deformation parameters for the $^{186,188,190}$Hg isotopes. 
The results are obtained with the Gogny-D1M interaction within the constrained Hartree-Fock-Bogoliubov approach and serve as a basis for the 5DCH calculations.}
\end{figure*}

In view of the measured lifetimes, many of them obtained for the first time, the neutron-deficient $^{186,188,190}$Hg have been studied within a self-consistent beyond-mean-field framework~\cite{PS_91_073003_2016, JPG_46_013001_2019}, i.e. the five-dimensional collective Hamiltonian (5DCH)~\cite{5DCH} and the symmetry-conserving configuration mixing (SCCM)~\cite{rodriguez2010triaxial, rodriguez2014structure} methods, with the Gogny-D1M~\cite{PhysRevLett.102.242501} and Gogny-D1S~\cite{PhysRevC.21.1568, BERGER1991365} interactions, respectively.
These calculations are based on the mixing of a set of intrinsic states with different quadrupole deformations. 


In the 5DCH approach, the intrinsic states are obtained by constrained Hartree-Fock Bogolyubov calculations (CHFB) performed for around 90 points in the $(\beta_2, \gamma)$ plane, defined by $0 \leq {\beta_2}_{max} \leq 0.8$ and $0^\circ \leq \gamma \leq 60^\circ$. 
These calculations are performed using a triaxial harmonic oscillator basis including 13 major shells, which is a sufficiently large model space to ensure the energy convergence.
The overlap of the intrinsic states is assumed to be gaussian: the Gaussian-overlap approximation (GOA) allows to derive from the Hill-Wheeler equation a Bohr-type Hamiltonian, called the 5DCH. 
This Hamiltonian deals with quadrupole degrees of freedom only, thus two vibrations and three rotations are taken into account. 
The 5DCH potential term is determined by the CHFB energy, obtained at the different $(\beta_2, \gamma)$ deformation points, to which the zero-point energies are added. 
The kinetic terms involving three mass parameters and three rotational inertia parameters are also deduced from CHFB solutions. 

\begin{figure*}[t]
\includegraphics[width=\textwidth]{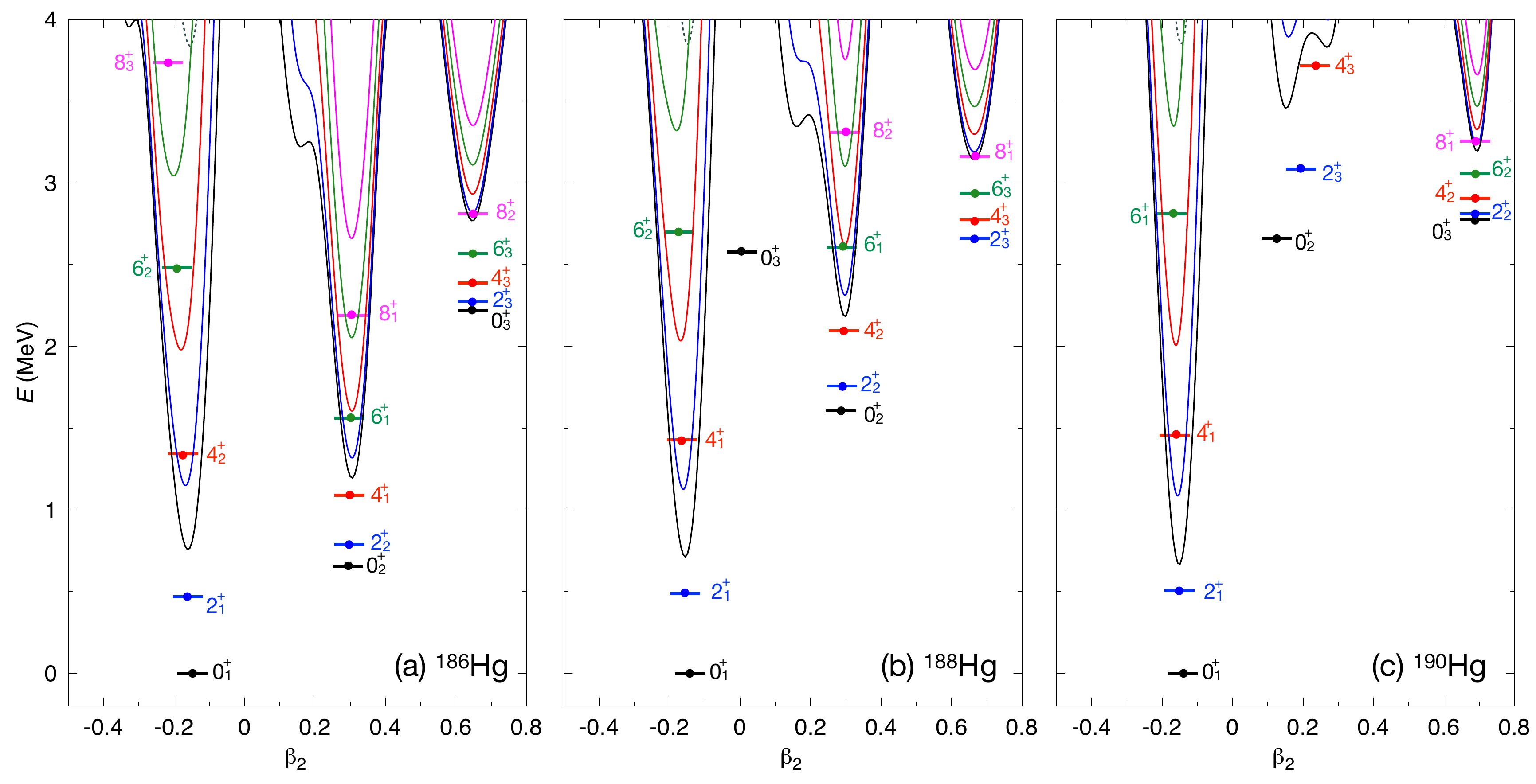}
\vspace{-8mm}
\caption{\label{fig:SCCM_SPECTRUM} (Color online) Energy spectra obtained with axial SCCM calculations with the Gogny-D1S interaction for the (a) $^{186}$Hg, (b) $^{188}$Hg and (c) $^{190}$Hg isotopes. 
The energies are plotted at the mean value of the intrinsic quadrupole deformation of the corresponding collective wave functions. 
PNAMP energy curves as a function of $\beta_{2}$ are also plotted to guide the eye.}
\end{figure*}


In the SCCM approach, the intrinsic states are Hartree-Fock-Bogoliubov (HFB) like wave functions obtained self-consistently through the particle-number variation after projection (PN-VAP) method~\cite{anguiano2001particle}. 
Because the HFB states also break the rotational invariance of the system, this symmetry is restored by projecting onto good angular momentum (particle-number and angular momentum projection, PNAMP). 
The final spectrum and nuclear wave functions are obtained by mixing such PNAMP states within the generator coordinate method (GCM). 
The region under study is expected to show competing shapes in the low-lying spectrum. 
Such kind of system can be particularly sensitive to the convergence of the calculations with the number of harmonic oscillator shells, $N_{h.o.}$, used to define the HFB transformation~\cite{PRC_91_044315_2015}. 

Initially the SCCM calculations were performed with $N_{h.o.}=11$ including triaxial shapes with the result that the prolate band was predicted to be the ground-state configuration, contrary to what is obtained experimentally. 
Consequently, exploratory SCCM calculations limited to the axial quadrupole ($\beta_{2}$) degree of freedom were performed for $N_{h.o.}= 11$, 13, 15 and 17. 
A strong dependence of the excitation energies of the less-deformed band was observed as a function of $N_{h.o.}$, with the inversion of the bands disappearing for larger bases. 
On the other hand, the $\beta_{2}$ deformation of the two structures was almost independent on the number of harmonic oscillator shells: $\beta_{2} \approx 0.20$ for the oblate-deformed configuration and $\beta_{2} \approx 0.25$ for the prolate one. 
Only for $N_{h.o.}=17$ convergence was obtained for the energies and wave functions.
Such a large number of harmonic oscillator shells makes the SCCM with triaxial shapes extremely expensive from the computational point of view and, consequently, only axial SCCM results with $N_{h.o.}=17$ are reported here.

\begin{figure*}[t]
\includegraphics[width=\textwidth]{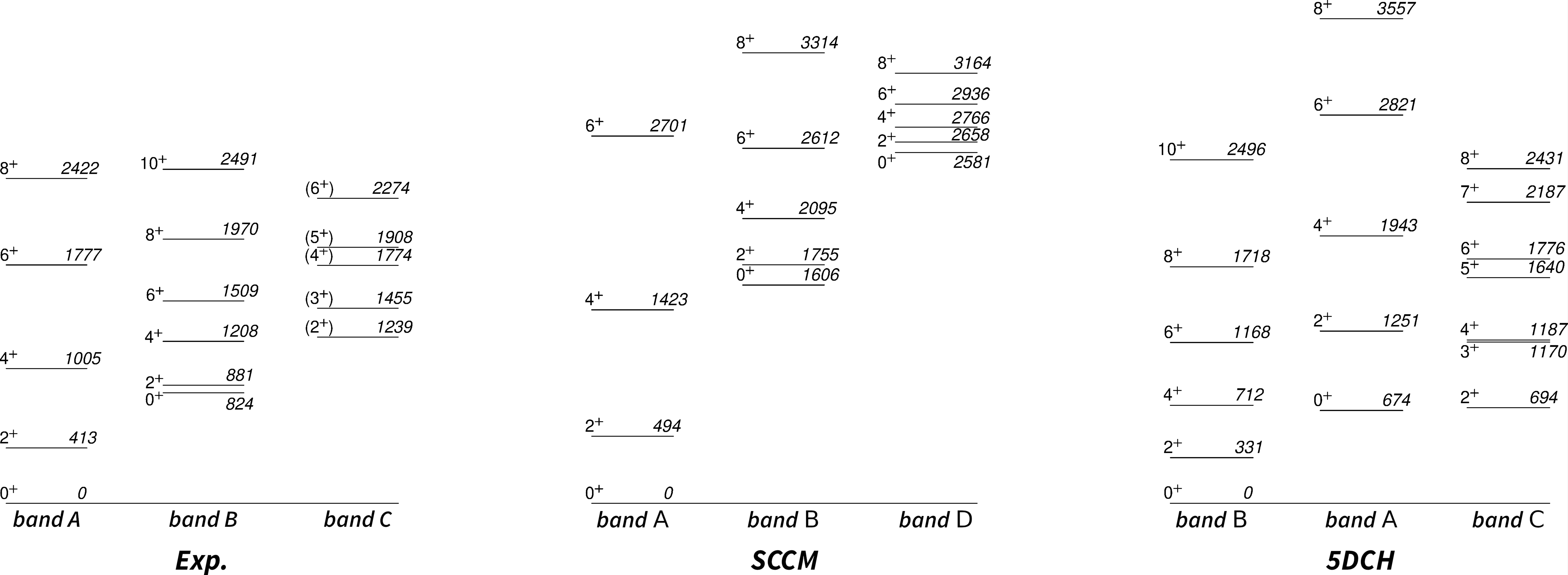}
\vspace{-5mm}
\caption{\label{fig:LevelScheme} (Color online) Partial level scheme of $^{188}$Hg, reporting the low-lying excited states. 
The experimental results are compared with the energy spectra predicted by the 5DCH and SCCM approaches.}
\end{figure*}

A first estimation of the structure of the $^{186,188,190}$Hg isotopes can be obtained by analysing the calculated energy surfaces as a function of deformation parameters. 
Figure~\ref{fig:SCCM_PES} presents the PN-VAP and PNAMP energies, calculated within the SCCM approach, as a function of the deformation $\beta_{2}$.
For each of the three nuclei two clear minima are present, corresponding to oblate ($\beta_{2}\approx-0.17$) and prolate superdeformed (SD) ($\beta_{2}\approx+0.65$) shapes, and a double minimum structure around normal-deformed (ND) prolate configurations with $\beta_{2}\approx+0.1,+0.3$.
Once the angular momentum projection is performed, in the three cases, the absolute minimum for $J=0$ is the oblate one, then the ND configuration  and, finally, the SD one. 
Oblate and ND minima are rather close in energy in $^{186}$Hg, while ND and SD minima are almost degenerate in $^{190}$Hg. 

Similar conclusions can be drawn from the analysis of the potential-energy surfaces (PES) for $^{186,188,190}$Hg (see Figure~\ref{fig:5DCH_PES}), serving as a basis for the 5DCH approach. 
For $^{188}$Hg, the oblate minimum with $\beta_2 \approx 0.16$ is the deepest one, followed by a slightly more shallow prolate-like deformed minimum with a significant degree of triaxiality $(\beta_2, \gamma)=(0.28, 13^\circ)$, and finally an axial prolate-superdeformed minimum with $\beta_2 \approx 0.65$, which is predicted much higher in energy. 
The PES for $^{186}$Hg is very similar, with the only difference that the two ND minima are almost degenerate.
On the contrary, for $^{190}$Hg the prolate-like normal-deformed minimum disappears, while the SD prolate minimum becomes more pronounced, in line with the conclusions of Ref.~\cite{DELAROCHE1989145}.
One can clearly conclude that the energy surfaces resulting from both calculations suggest the presence of shape coexistence in these nuclei. 

As already mentioned, the final theoretical spectra are obtained by mixing intrinsic HFB-like states having different quadrupole deformations.
In the SCCM calculations, the GCM is applied to mix the PNAMP states. 
In the 5DCH approach, the intrinsic CHFB states are mixed assuming a gaussian overlap to solve the GCM-equivalent equation. 
Figure~\ref{fig:SCCM_SPECTRUM} shows the excitation energies, obtained with the SCCM approach, for the $^{186,188,190}$Hg nuclei as a function of the mean value of the intrinsic quadrupole deformation for each state provided by the collective wave functions (CWF). 
Here, we clearly observe for the three isotopes the three different collective bands (with $\Delta J=2$) associated to the oblate (ground-state bands), ND-prolate and SD-prolate minima. 
The excitation energy of the $0^{+}_{2}$ state in $^{186}$Hg is lower than those for $^{188,190}$Hg, as observed experimentally. 
For $^{190}$Hg, the $0^{+}_{2}$ state is found close to the band-head of the SD band. 

Figure~\ref{fig:LevelScheme} presents a comparison of the partial level scheme of $^{188}$Hg with the theoretical spectra, calculated within the SCCM and 5DCH approaches. 
Although the absolute minimum of the PES in Figure~\ref{fig:5DCH_PES} is located at an oblate deformation, the oblate-like (\textit{band A}) and prolate-like (\textit{band B}) structures resulting from the 5DCH calculations are inverted with respect to experimental data. 
Nevertheless, the energy difference between the two $0^+$ states is 674 keV, which is similar to what is observed
experimentally, and suggests that mixing between the two structures may be correctly reproduced. 
In contrast, while the SCCM calculation reproduces the proper ordering of the bands, it predicts a much larger excitation energy of the $0_2^+$ state, which can be related to the absence of the triaxial degree of freedom
and result in a smaller mixing between the excited states. 
At the same time, the SCCM calculation predicts a band crossing at $J^{\pi}=6^{+}$ in $^{188}$Hg in agreement with the observations, i.e. the members of the ND-prolate band become yrast at $J=6$, 8, 10. 
Correct band crossing is also predicted for $^{186,190}$Hg.
Regarding the level spacing of the two bands, the SCCM method reproduces better the properties of the ND-prolate band (\textit{band B}), while the 5DCH approach provides a better description of the weakly deformed structure (\textit{band A}). 
Due to the included triaxial degree of freedom, the 5DCH is able to predict a $\gamma$-band (\textit{band C}) appearing at low excitation energy. 
Such a structure has not been observed in the experiments presented in this work, but it was first postulated in a $\beta$-decay study of Ref.~\cite{bourgeois1976hg} and then confirmed in Ref.~\cite{PhysRevC.30.1267}, although the existing spin assignments are not firm. 

\begin{figure}[t]
\includegraphics[width=0.78\columnwidth]{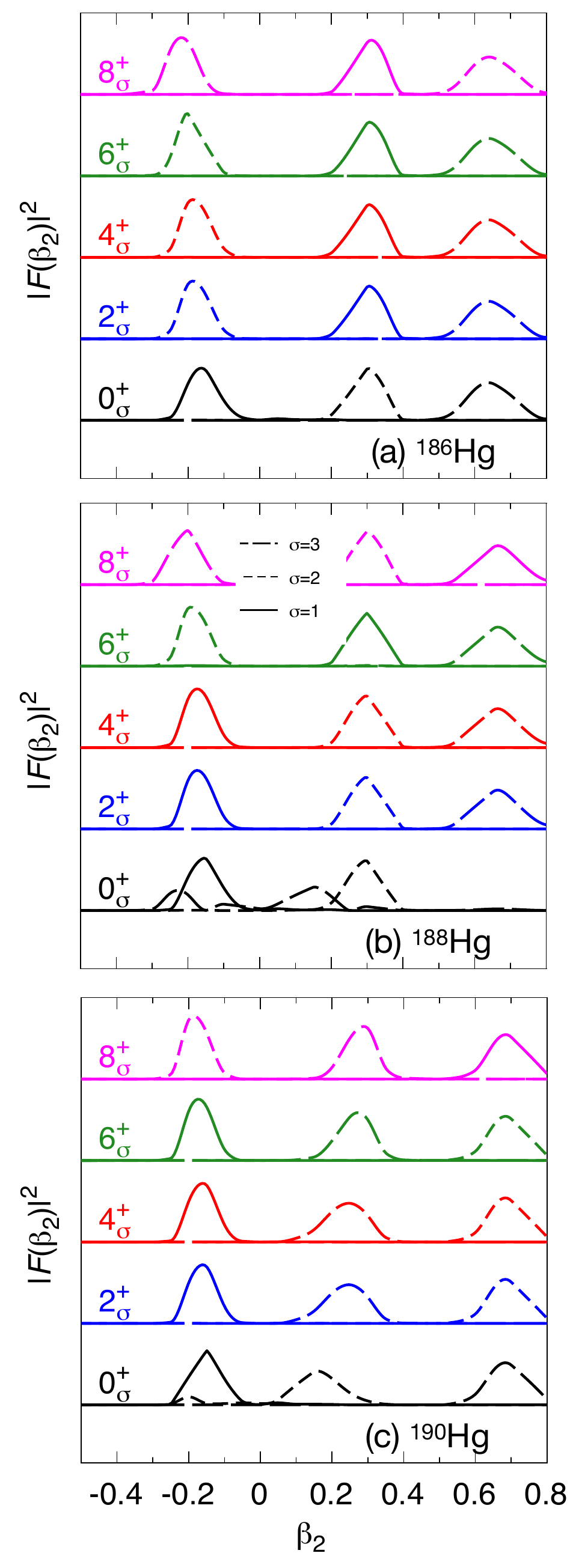}
\vspace{-5mm}
\caption{\label{fig:SCCM_COLL_WF} (Color online) Collective wave functions (CWF) obtained with axial SCCM calculations with the Gogny-D1S interaction for (a) $^{186}$Hg, (b) $^{188}$Hg and (c) $^{190}$Hg isotopes.}
\end{figure}
\begin{figure}[t]
\centering
\includegraphics[width=\columnwidth]{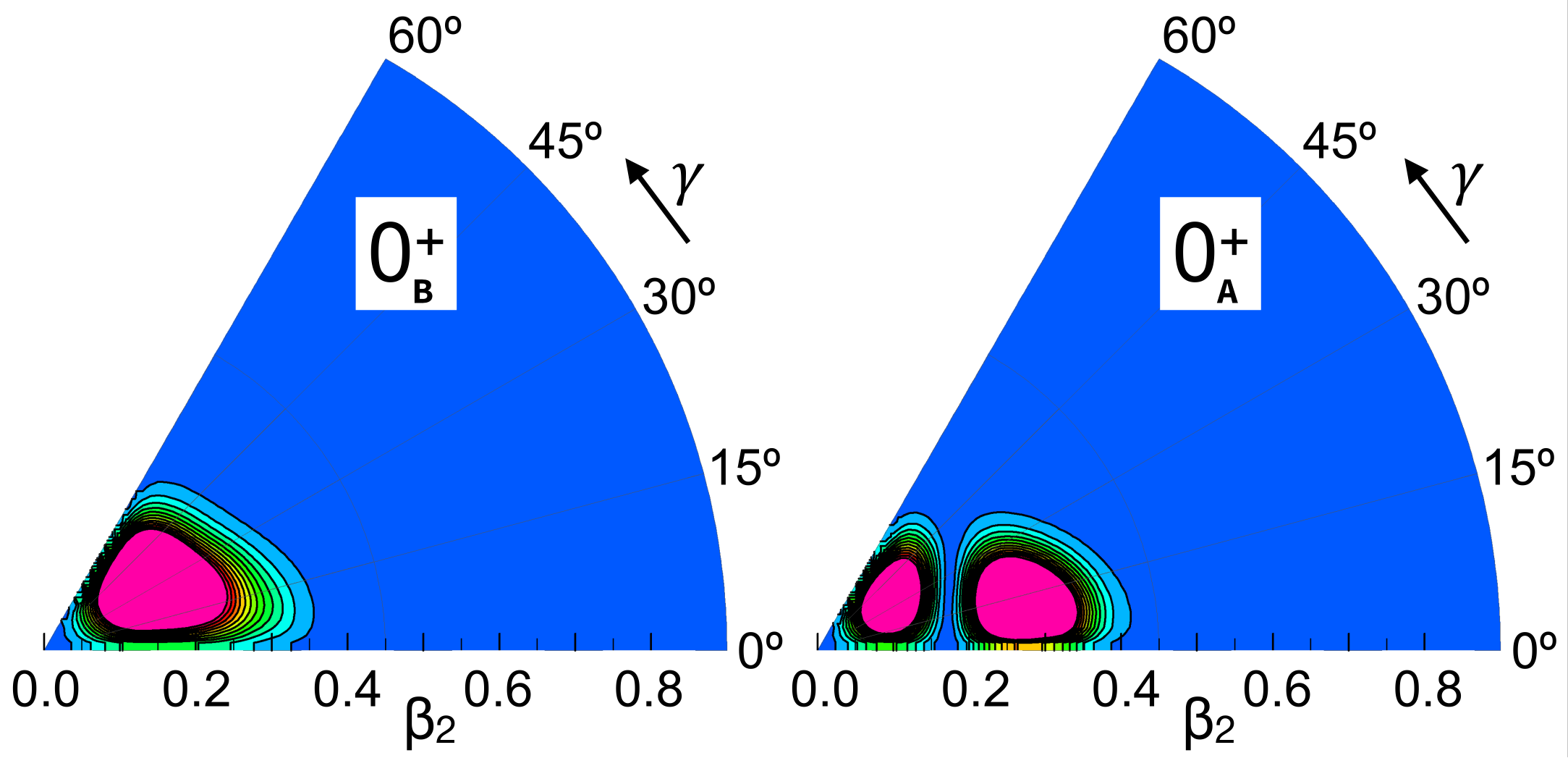}
\vspace{-6mm}
\caption{\label{fig:5DCH_COLL_WF} (Color online) Collective wave functions (CWF) as a function of the
$\beta_{2}$ and $\gamma$ deformation parameters for the $0_A^+$ and $0_B^+$ states, obtained with the Gogny-D1M interaction within the 5DCH framework.}
\end{figure}

Finally, a potential energy minimum corresponding to a superdeformed shape is predicted in both theoretical approaches. 
In this work, the superdeformed states were only obtained in SCCM calculations (\textit{band D}). 
Indeed, in the 5DCH approach a two-center basis has to be adopted for diagonalising the Bohr Hamiltonian in order to explore the highly deformed structure~\cite{DELAROCHE1989145}. 
Such a structure has not been observed in $^{188}$Hg, while it is known in the neighboring $^{189}$Hg and $^{190}$Hg~\cite{DRIGERT1991452, Bearden1992, BEARDEN1994441}.

The CWF, i.e. the weights of the intrinsic quadrupole deformations in each nuclear state, are presented in Figure~\ref{fig:SCCM_COLL_WF} and Figure~\ref{fig:5DCH_COLL_WF} for the SCCM and 5DCH methods, respectively.
The SCCM calculations predict a rather constant deformation within each of the three calculated bands (oblate \textit{band A} with $\beta_{2}\approx -0.18$, normal-deformed prolate \textit{band B} with $\beta_{2}\approx 0.22$, and superdeformed \textit{band D} with $\beta_{2}\approx 0.65$), with the only exception of the $0^+_3$ state being much less deformed than other members of the SD band. 
For the ground-state and ND bands, the deformation parameters from the SCCM calculations are in fair agreement with those estimated from the experimental results, assuming the rotational model (see Table~\ref{tab:B(E2)}).
The overlaps between the CWFs are negligible in most cases, which can be related to a very small mixing between the three structures and attributed to the imposition of axial symmetry in the calculations.
An exploratory calculation with $N_{h.o}=11$ including triaxial shapes for $^{188}$Hg suggests a larger mixing between the oblate-like and ND prolate-like configurations along the $\gamma$ degree of freedom.
In the axial SCCM calculation noticeable overlaps of CWFs can be observed only for the 0$^+$ states, which suggests that the three structures (oblate, prolate and SD) mix.

The wave functions resulting from the 5DCH approach are much more spread on the $(\beta_2,\,\gamma)$ plane and the distributions calculated for states belonging to the different structures present considerable overlaps, see Figure~\ref{fig:5DCH_COLL_WF}. 
Due to the importance of the triaxial degree of freedom, it is impossible to refer to the two bands as ``prolate'' and ``oblate''. 
The average $\beta_2$ deformation of states belonging to the \textit{band B}, corresponding to the experimentally observed deformed band built on the $0^+_2$ state in $^{188}$Hg, slowly increases with spin starting from $\beta_2=0.2$ for the $0^+$ state, which is in fair agreement with the $\beta_2$ deformation estimated from the experimental data. 
A gradual evolution from an almost maximally triaxial shape with $\gamma=27^\circ$ for the $0^+$ state towards a more prolate deformation higher up in the band is also predicted. 
Notably, the average $\gamma$ deformation parameter for the states of spin $J\ge4$ is about $15^\circ$, very close to the value derived from the experimental transition probabilities using the asymmetric rotor model (see Section IV). 
Interestingly, an average $\gamma$ deformation parameter of about 15$^\circ$ was also obtained for the states belonging to \textit{band B} in the exploratory SCCM calculations including the triaxial degree of freedom, but limited to $N_{h.o}=11$. 
The average elongation parameter of states belonging to \textit{band A}, corresponding to the experimental ground-state band in $^{188}$Hg, is $\beta_2 \approx 0.25$ and remains constant within the band, while the triaxial parameter evolves from about $\gamma \approx 20^\circ$ towards maximum triaxiality. 
The states belonging to this oblate-like structure present strong admixtures of the K=2 configuration, being as large as almost $50\%$ for the $2_A^+$ state.

Both calculations yield a similar picture of two structures with $|\beta_2| \approx 0.2$, coexisting at low excitation energy. 
The 5DCH calculations suggest that at low-spin the triaxial degree of freedom is important. 
Unfortunately, the precision of the experimentally determined quadrupole moment of the 2$^+_1$ state in $^{188}$Hg, equal to 0.8$^{+0.5}_{-0.3}$ eb~\cite{wrzosek2019coulex} is insufficient to make a conclusion about  the degree of triaxiality in this state. 
This value is compatible with both calculations: the SSCM calculations give $Q_s(2^+_A) = 1.3$ eb, while in the 5DCH calculations, which attribute a triaxial character to the $2^+_A$ state, the spectroscopic quadrupole moment is $Q_s(2^+_A) = 0.5$ eb.


Table~\ref{tab:B(E2)} presents a comparison of the experimental reduced transition probabilities in $^{188}$Hg with those calculated within the SCCM and 5DCH approaches. 
In order to facilitate the comparison, the 5DCH results for the ground-state band are reported next to those for the more deformed structure (\textit{band A}), while those for the structure built on the 0$^+_2$ are listed next to those for the experimental weakly deformed ground-state band (\textit{band B}).

\begin{table*}[t]
\caption{\label{tab:mixing}Average quadrupole moments (in $e$b) of the normal and intruder configurations as well as the mixing amplitudes $\alpha_{mix}$ for the $0^+_{g.s.}$, $2^+_1$ and $4^+_1$ states extracted from the theoretical $B(E2)$ values from 5DCH and SCCM calculations using the two-band mixing model, compared with those deduced from experimental data (see text). 
Due to the bands swapping, for the 5DCH and SCCM ($N_{h.o.}=11$) calculations the values are interchanged to facilitate a direct comparison with the experimental results.}
\centering
\begin{ruledtabular}
\begin{tabular}{c | c c c | c c}
		&	$\alpha_{mix}(0^+_{g.s.})$	& $\alpha_{mix}(2^+_1)$	&	$\alpha_{mix}(4^+_1)$	&	$|Q_0^n|$	&	$|Q_0^i|$	\\
\hline
Exp.					&	0.995		&	0.988		&	0.893	&	3.66(37)	&	6.90(57)\\
5DCH					&	0.720		&	0.866		&	0.942	&	4.20	&	7.68	\\
SCCM ($N_{h.o.}=11$)	&	0.800		&	0.911		&	0.949	&	3.12	&	9.61	\\
SCCM ($N_{h.o.}=17$)	&	0.993		&	0.980		&	0.994	&	4.45	&	8.45	\\
\end{tabular}
\end{ruledtabular}
\end{table*}

The agreement for in-band transitions is reasonably good, with the transitions within the weakly deformed band (\textit{band A}) described better by SCCM, while those in the more deformed structure (\textit{band B}) are overestimated by both models, with the 5DCH results being closer to experimental values. 
The $B(E2)$ value for the $6_1^+ \to 4_1^+$ inter-band transition is underestimated by both calculations, 
with the one resulting from the SCCM calculations being more than one order of magnitude smaller than the experimental value. 
This can be interpreted as an underestimation of the possible mixing between the two bands, consistent with the negligible overlaps between the CWFs (Figure~\ref{fig:SCCM_COLL_WF}).
For a more quantitative comparison,
one can extract the mixing amplitudes $\alpha_{mix}$ (see Equation~\ref{eq-mixing}) for the $0^+$, $2^+$ and $4^+$ states from the calculated B(E2) transition probabilities using the equations listed in Section V.A of Ref.~\cite{clement2007kr}.
The values obtained using this approach from the experimental data and from the model calculations are reported in Table~\ref{tab:mixing}. 
As expected, the mixing strengths fo r the $0^+$ and  $2^+$ states obtained within the 5DCH approach are much larger than for the SCCM calculations, and also largely exceeding what is observed experimentally. 
In contrast, the large mixing strength for the $4^+$ states deduced from the experimental data is better reproduced by the 5DCH calculations. 

\begin{figure}[t]
\includegraphics[width=0.87\columnwidth]{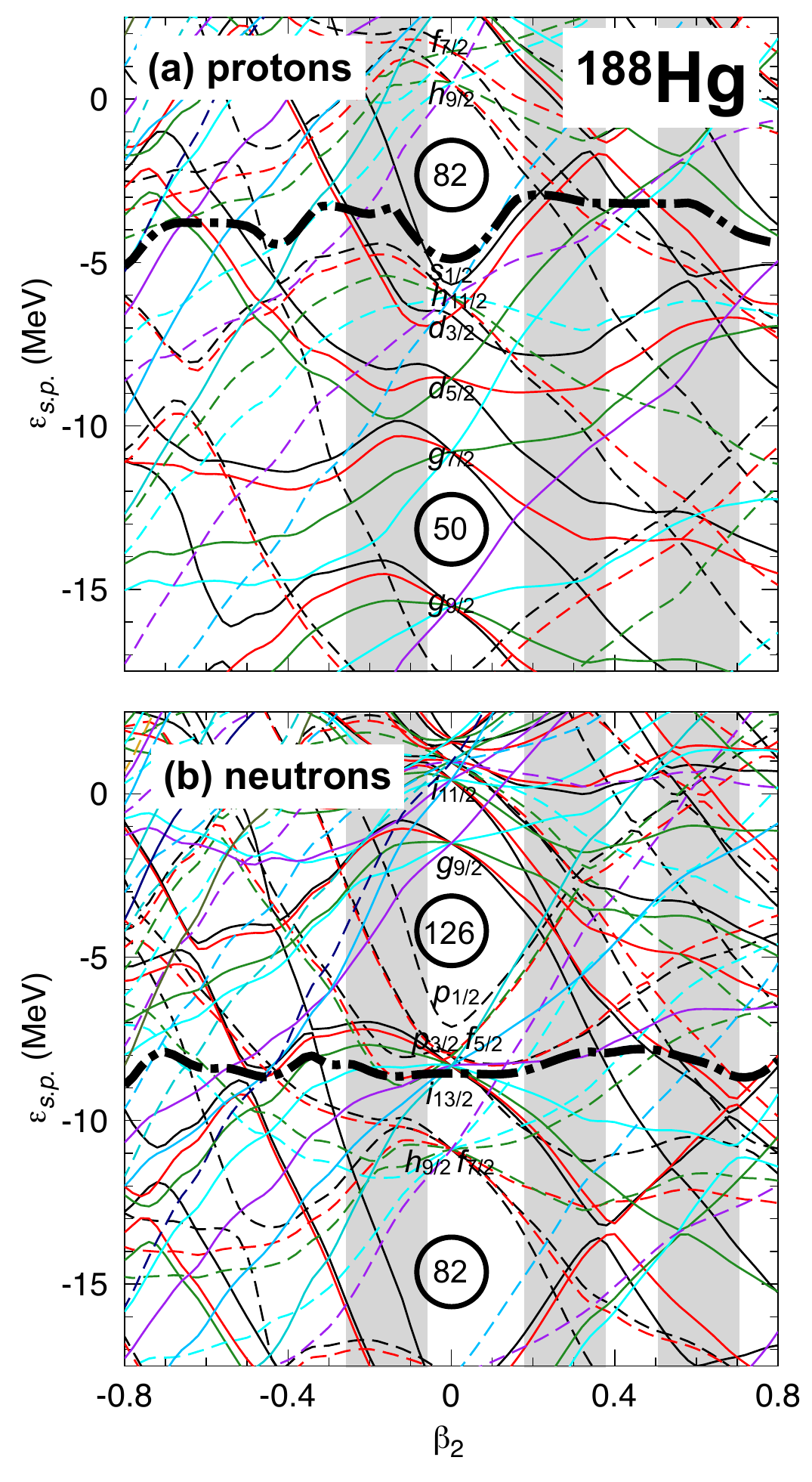}
\vspace{-5mm}
\caption{\label{fig:SCCM_Nilsson} (Color online) Single-particle energies for (a) protons and (b) neutrons as a function of the deformation $\beta_{2}$ calculated for $^{188}$Hg with the Gogny D1S interaction. 
Dashed (continuous) lines represent negative (positive) parity states and the thick dashed-dotted lines show the Fermi energy. 
The shaded areas correspond to the position of the three minima in the energy surface shown in Figure~\ref{fig:SCCM_PES}.}
\end{figure}

Finally, the underlying shell structure of the three collective bands can be also analysed within the SCCM framework. 
The relevant spherical shells that are needed to describe such states can be qualitatively identified by Nilsson-like orbitals. 
These orbitals are computed self-consistently for each nucleus with HFB states obtained along the quadrupole degree of freedom. 
As an example, the evolution of the  orbitals as a function of the deformation, $\beta_{2}$, is plotted for the $^{188}$Hg isotope in Figure~\ref{fig:SCCM_Nilsson}. 
The Fermi energies for protons and neutrons are also shown. 
The relevant regions, where the CWF are peaked (i.e. the minima of the energy surfaces), are marked by a grey band in the figure. 
In the oblate minimum, we see that the orbitals close to the Fermi energy are: $3s_{1/2}$, $0h_{11/2}$ and $0h_{9/2}$ for protons and $0h_{9/2}$, $1f_{7/2}$, $0i_{13/2}$, $2p_{3/2}$ and $1f_{5/2}$ for neutrons. 
For the ND configuration, the $1d_{5/2}$ and the $1f_{7/2}$ for protons and the $0g_{9/2}$ for neutrons also play a role. 
Finally, for the SD band, the $0i_{11/2}$ and negative parity orbitals above $N=126$ are also crossing the neutron Fermi energy. 
This picture shows the complexity of the single-particle structure of these heavy nuclei that prevents the use of shell model calculations. 
We compute the occupation numbers of spherical orbitals with each individual SCCM wave function in the three isotopes studied here following the method developed in Ref.~\cite{PRC_93_054316_2016}. 
To simplify the discussion, we define a proton core of $Z=82$ that encompasses the $0s$, $0p$, $1s-0d$, $1p-0f$, $2s-1d-0g$ and $0h_{11/2}$ spherical orbitals, and a neutron core of $N=100$ that includes the orbitals listed
above plus the $0h_{9/2}$ and $1f_{7/2}$ neutron levels. 
Using such cores, the $^{188}$Hg isotope in a normal filling approximation should correspond to 2h-0p (protons) and 0h-8p (neutrons), respectively. 
However, the results obtained for the oblate, ND and SD bands are 4.0h-2.0p, 8.1h-6.1p and 9.0h-7.0p for protons, and 5.4h-13.4p, 7.7h-15.7p and 8.3h-16.3p for neutrons, respectively. 
Similar configurations are also obtained for the other two isotopes calculated here. 
These results show again the complexity of the single-particle structure driven by the deformation of the system.

\section{Conclusions}

The nuclear structure of the neutron-deficient mercury isotope $^{188}$Hg  was investigated via lifetime measurements at the Laboratori Nazionali di Legnaro.
In order to control a possible contamination of the $^{188}$Hg data by other reaction channels, two different fusion-evaporation reactions were used for this study and, thanks to the powerful capabilities of both GALILEO and Neutron-Wall arrays, a clear identification of the channel of interest was possible. 

Using the RDDS technique, the lifetimes of the states up to spin $16 \, \hbar$ were measured 
and the results obtained for the $2_1^+$ and $4_1^+$ states are in agreement with the values reported in literature.  
Thanks to the new results for both low- and high-lying states, the deformation of the ground-state band and of the intruder structure was estimated from the experimental data, assuming two-band mixing and two different rotational models. 
These models provide a similar interpretation of the structures: the ground-state band has a quadrupole deformation of $\beta_2 \approx 0.13$, while the intruder one has $\beta_2 \approx 0.22$.
Moreover, the lifetimes of the $14_1^+$ and $16_1^+$ excited states highlighted the presence of an almost spherical structure above the $12_1^+$ isomer. 

In view of the new results, two state-of-the-art beyond-mean-field calculations were performed for the even-mass $^{186-190}$Hg nuclei using the symmetry-conserving configuration-mixing approach limited to axial shapes and the 5-dimensional collective Hamiltonian including the triaxial degree of freedom. 
Both calculations yield a similar picture of two structures with $|\beta_2| \approx 0.2$, coexisting at low excitation energy, but predict different relative positions of the two bands and their mixing.
The underlying shell structure of the collective bands was analysed within the SCCM framework, identifying the relevant spherical shells, necessary to describe such structures. 
For $^{188}$Hg the comparison between the theoretical predictions and the experimental results confirmed the proposed interpretation of both the intruder and ground-state bands. 

In contrast with other beyond-mean-field calculations, the SCCM approach predicts the presence of shape coexistence also in $^{190}$Hg, with the bandhead of the SD structure mixing with those of the ND prolate and oblate bands. 
Clearly, further experimental studies of $^{190}$Hg are required to verify these predictions.
The mixing amplitudes for the 0$^+$ states in $^{180-188}$Hg deduced from level energies and E2 transition strengths differ from those obtained from $\alpha$-decay hindrance factors. 
In this context, measurements of  $\rho^2(E0)$ between the coexisting structures are called for, as they would provide a direct measure of the degree of their mixing.

\section*{Acknowledgements}
\vspace*{-3mm}
The authors would like to thank the GALILEO collaboration. 
Special thanks go to the INFN-LNL technical staff for their help and the good quality beam and to Massimo Loriggiola for the extended work on the Gd plunger targets.
We acknowledge the numerical calculations of S. Hilaire.
This paper owes much to the collaboration with P.E. Garrett. 
This work was partially supported by the Espace de Structure Nucl\'{e}aire Th\'{e}orique (CEA/DSM-DAM).
The work of T.R.R. was supported by the Spanish MICINN under grant no. PGC2018-094583-B-I00.
The work was also supported (P.R.J. and P.K.) by the German BMBF under contract no. 05P18RDFN9, (G.J.) by the National Science Centre Poland (NCN) under the grant no. 2017/25/B/ST2/01569.
A.G. is grateful to the support of Fondazione Cassa di Risparmio Padova e Rovigo.    

\bibliography{ShapeHg}
\bibliographystyle{epj}

\end{document}